\documentclass[11pt,a4paper]{article}
\pdfoutput=1
\usepackage{jcappub}
\usepackage[utf8]{inputenc}
\usepackage[normalem]{ulem}
\usepackage{tikz,xcolor}
\usepackage{pgf}
\usetikzlibrary{decorations.pathreplacing}
\usetikzlibrary{arrows,shapes}
\usepackage{amsfonts}
\usepackage{fullpage}
\usepackage{caption}
\usepackage{subcaption}

\newcommand{\kv}{\mathbf{k}}
\newcommand{\qv}{\mathbf{q}}

\newcommand{\pv}{\mathbf{p}}

\newcommand{\fnl}{f_{\rm NL}}
\newcommand{\gnl}{g_{\rm NL}}
\newcommand{\zg}{\zeta_{\rm G}}
\newcommand{\zng}{\zeta_{\rm NG}}
\newcommand{\benv}[1]{\begin{#1}}
\newcommand{\eenv}[1]{\end{#1}}

\newcommand{\be}{\benv{equation}}
\newcommand{\ee}{\eenv{equation}}
\newcommand{\nn}{\nonumber}
\newcommand{\fref}[1]{Fig.\,\ref{#1}}

\def\dthree#1{\frac{d^3{#1}}{(2\pi)^3}}

\title{Cosmic variance in inflation with two light scalars}
\author[1]{B\'eatrice Bonga,}\emailAdd{bpb165@psu.edu}
\author[1]{Suddhasattwa Brahma,}\emailAdd{suddhasattwa.brahma@gmail.com}
\author[1]{Anne-Sylvie Deutsch}\emailAdd{asdeutsch@psu.edu}
\author[1,2]{and Sarah Shandera}\emailAdd{shandera@gravity.psu.edu}

\affiliation[1]{Institute for Gravitation and the Cosmos \& Physics Department, The Pennsylvania State University, University Park, PA 16802, USA}
\affiliation[2]{Perimeter Institute for Theoretical Physics,  31 Caroline Street North, Waterloo, Ontario N2L 2Y5 Canada}

\abstract{We examine the squeezed limit of the bispectrum when a light scalar with arbitrary non-derivative self-interactions is coupled to the inflaton. We find that when the hidden sector scalar is sufficiently light ($m\lesssim0.1\,H$), the coupling between long and short wavelength modes from the series of higher order correlation functions (from arbitrary order contact diagrams) causes the statistics of the fluctuations to vary in sub-volumes. This means that observations of primordial non-Gaussianity cannot be used to uniquely reconstruct the potential of the hidden field. However, the local bispectrum induced by mode-coupling from these diagrams always has the same squeezed limit, so the field's locally determined mass is not affected by this cosmic variance.}

\begin{document}
\maketitle
\flushbottom


\section{Introduction}
The statistics of the primordial fluctuations beyond the power spectrum contain information about the spectrum and interactions of light particles present during inflation. Although the {\it Planck} satellite bounds on non-Gaussianity are excellent ($\sigma(\fnl)\sim\mathcal{O}(10)$ \cite{Ade:2015ava}), they do not yet cross even the highest theoretically interesting (and roughly shape-independent \cite{Alvarez:2014vva}) threshold to rule out $\fnl\sim\mathcal{O}(1)$. Future data will help probe the remaining parameter space, and either stronger constraints or detection of non-Gaussianity would provide an important clue about physics at the inflationary scale.

When the primordial fluctuations are entirely or partly sourced by a light field other than the inflaton, the correlation functions can have the interesting property that locally measured statistics depend on the realization of long wavelength modes. For example, the ``local" bispectrum generated in curvaton scenarios \cite{Linde:1996gt, Moroi:2001ct, Lyth2001, Enqvist:2001zp, Dvali:2003em,Zaldarriaga:2003my} correlates the amplitude of short wavelength fluctuations with the amplitudes of much longer wavelength modes. This means that in the presence of local ansatz non-Gaussianity, the power spectrum measured in sub-volumes will vary depending on how much the background fluctuations deviate from the mean \cite{Nelson:2012sb,Nurmi:2013xv, LoVerde:2013xka,LoVerde:2013dgp}. While the bispectrum can cause the power spectrum to vary in sub-volumes, the trispectrum can generate shifts to the bispectrum and the power spectrum in biased sub-volumes (that is, sub-volumes whose long wavelength background is not the mean). More generally, long wavelength modes in the $n$-point function can shift any lower order correlation function \cite{Nelson:2012sb, Byrnes:2013ysj}. Since nearly all sub-volumes will have a long wavelength background that is not zero, statistics measured in any small region are likely to be biased compared to the mean statistics of the larger volume.

This correlation between local statistics and the long wavelength background can be used to detect non-Gaussianity (e.g., through the non-Gaussian halo bias \cite{Dalal:2007cu}) when applied to sub-volumes where at least some long wavelength modes are observable. But it may also be relevant to the conclusions we can draw about inflationary physics: our entire observable universe is almost certainly a sub-volume of a larger, unobservable, space. If the perturbations we observe turn out to have any form of long-short mode coupling we must assume there is a `super cosmic variance' uncertainty in comparing observations in our Hubble volume with the mean predictions of inflation models with more than the minimum number of e-folds \cite{Bartolo:2012sd,Nelson:2012sb,Nurmi:2013xv, LoVerde:2013xka,Thorsrud:2013mma, LoVerde:2013dgp, Thorsrud:2013kya, Byrnes:2013ysj, Bramante:2013}. In that case, there is not necessarily a one-to-one map between properties of the correlation functions we measure and parameters in an inflationary Lagrangian.  In this paper we will explore an example where some, but not all, of the parameters of the Lagrangian are obscured by cosmic variance.

Mathematically, not all non-Gaussian fields have statistics that differ in sub-volumes. In inflationary models the influence of long wavelength fluctuations on locally observed correlations depends on how many degrees of freedom source the background expansion and the observed fluctuations. Single-clock inflation has no significant coupling between modes of very different wavelengths \cite{Maldacena:2002vr, Creminelli:2004yq, Hinterbichler:2013dpa, Joyce:2014aqa, Mirbabayi:2014zpa}: on a suitably defined spatial slice, the statistics in any sub-volume are the statistics of the mean regardless of the level of non-Gaussianity or the amplitudes of long wavelength fluctuations. In contrast, the family of non-Gaussian fields built from arbitrary local functions of a Gaussian field couples short wavelength modes to all long wavelength modes. As a result, the amplitude of fluctuations and the amplitude of non-Gaussianity ($\fnl^{\rm local}$) can vary by large factors in sub-volumes \cite{Nelson:2012sb,Nurmi:2013xv, LoVerde:2013xka,Bramante:2013}. In fact, regardless of whether the mean statistics are weakly or strongly non-Gaussian, sub-volumes that are sufficiently biased (that is, whose long wavelength background is sufficiently far from the mean) all have the statistics of a weakly non-Gaussian local ansatz. However, some properties of the correlation functions are the same in all volumes. In particular, the squeezed limit scaling of the bispectrum is always (very nearly) that of the local template \cite{Nelson:2012sb, LoVerde:2013xka}. The preservation of the shape of the bispectrum is not a generic feature of non-Gaussianity; it is easy to construct examples of non-Gaussian fields where the shape of the bispectrum (even in the squeezed limit) changes significantly in biased sub-volumes \cite{2015PhRvD..91h3518B}. Multi-field inflation models can produce a wide range of correlation functions, so it is worthwhile to understand more generally which properties of any potentially observable primordial non-Gaussianity are independent of the long wavelength background and which are not.

An interesting example that naturally interpolates between the single-clock and curvaton (local ansatz) mode coupling is quasi-single field inflation \cite{Chen:2009zp, Chen2010a}, where an additional scalar field is coupled to the inflaton during inflation. This field does not contribute to the background expansion and its self-interactions are not restricted by the approximate shift symmetry that the inflaton is subject to. Observational evidence for this `hidden sector' field would be found in the non-Gaussianity it indirectly sources in the adiabatic mode. Previous studies have shown that when the hidden sector field has a cubic self-interaction, the degree of long-short coupling in the observed bispectrum is determined by the mass of the second field (with the strongest coupling coming from a massless field). The coupling between modes of very different wavelengths is captured by the squeezed limit of the bispectrum, where one momenta corresponds to a much longer wavelength than the others (e.g, $k_1\ll k_2\approx k_3$). Measuring the dependence on the long wavelength mode ($k_1\equiv k_L$) in the squeezed limit of the bispectrum would reveal the mass of this spectator field \cite{Chen2010a, Chen2010b, Sefusatti:2012ye, Norena:2012yi} (see Eq.(\ref{eq:bisp-in-squeezed-limit}) below). The sensitivity of squeezed limits to the mass and spin of fields in more general multi-field scenarios have been discussed in \cite{Baumann2011b, Mirbabayi:2015hva, Arkani-Hamed:2015bza}.

In this paper we investigate how robust the quasi-single field bispectrum shape is to cosmic variance when higher order correlations are included in the model. When the hidden sector field is sufficiently light, any additional correlations may bias the statistics observed in sub-volumes. As a simple first case, we consider the correlations generated by contact diagrams from additional (i.e., quartic and beyond) non-derivative self-interactions of the hidden sector fluctuations. By computing the power spectrum and bispectrum in sub-volumes with non-zero long wavelength background modes we will show that (similarly to the local model), {\it any} non-derivative self-interaction of the spectator field leads to the same pattern of correlation functions in sufficiently biased sub-volumes. In particular, non-zero long wavelength fluctuations induce a tree-level bispectrum locally even if the mean theory does not contain one. Furthermore, the squeezed limit scaling of the bispectrum is the same in all sub-volumes, while the local amplitude of fluctuations and amplitude of non-Gaussianity are subject to cosmic variance. So, although the diagrams we consider here generate cosmic variance that obscures the details of a light hidden sector field's potential, that cosmic variance does not affect the measurement of the field's mass from squeezed limit of the locally observed bispectrum.

In the next section we review the quasi-single field scenario and compute some properties of arbitrary order correlation functions from additional self-interactions of the hidden sector scalar. In Section \ref{sec:shiftBS} we compute the statistics observed in sub-volumes. We discuss the results in Section \ref{sec:discussion}. The appendices contain some details of the calculations.


\section{Non-Gaussianity from quasi-single field inflation with \texorpdfstring{$\sigma^n$}{sigma to the n} interactions}
\label{sec:QSF}
Quasi-single field inflation \cite{Chen2010a} involves two fields: the inflaton $\Phi$ and a scalar $\Sigma$, coupled to the inflaton, whose energy density does not significantly contribute to the background expansion. The Lagrangian for the coupled perturbations, $\varphi$ and $\sigma$ respectively, is
\begin{equation}
	\mathcal{L} = -\frac{1}{2} (\partial \varphi )^2 + \rho \dot{\varphi} \sigma  - \frac{1}{2} (\partial \sigma)^2 - \frac{1}{2}m^2 \sigma^2 - V(\sigma).
	\label{eq:Lagrangian-QSFI}	
\end{equation}
Here $\varphi$ is related (at first order) to the curvature mode $\zeta$ by
\begin{equation}
\varphi = - \sqrt{2\epsilon} \zeta,
\label{eq:phi-to-zeta}
\end{equation}
where $\epsilon = -\dot{H}/H^2$ describes the evolution of the Hubble parameter $H$ (the dot denotes a cosmological time derivative). We have explicitly written out all quadratic terms in Eq.(\ref{eq:Lagrangian-QSFI}), so the potential $V(\sigma)$ starts at cubic order. Since $\Sigma$ is not the inflaton field, its interactions are not restricted by an approximate shift symmetry. In general the mass of the $\sigma$ fluctuation and the potential $V(\sigma)$ will depend on the original potential for $\Sigma$, expanded about some constant background vacuum expectation value $\Sigma_0$, as well as the terms coupling $\Sigma$ and $\Phi$.

\begin{figure}
\begin{center}
\begin{tikzpicture}
	\draw[thick] (0,0) -- (1,0);
	\draw[thick,dashed] (1,0) -- (2,0);
	\draw (1.2,.3) node{$\rho$};
	\draw[thick,dashed] (5,0) -- (6,0);
	\draw[thick,dashed] (6,0) -- (6.5,0.87);
	\draw[thick,dashed] (6,0) -- (6.5,-0.87);
	\draw (5.8,.3) node{$\mu$};
\end{tikzpicture}
\end{center}
\caption[Interaction terms for QSFI]{Interaction terms in quasi-single field inflation: a transfer vertex between $\varphi$ (solid line) and $\sigma$ (dashed line) with dimensionless coupling strength $\rho/H$, and the cubic self-interaction of~$\sigma$ with dimensionless coupling $\mu/H$.}
\label{fig:QSFI-interaction-terms}
\end{figure}
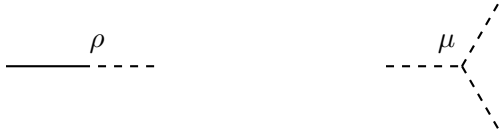

The coupling between the fields allows the curvature perturbation to inherit non-Gaussianity from the $\sigma$ self-interactions. Here we are only interested in the weak mixing case, that is, $\rho \ll H$, since this results in coupling between long and short modes. (In contrast, in the strong mixing case, the two fields effectively behave as a single degree of freedom with a modified speed of sound with no significant coupling between modes of different wavelengths~\cite{Baumann2011}.)
Although the quasi-single field model was first introduced with a very specific potential for both the adiabatic and isocurvature perturbations, with the inflaton field following a turning trajectory with a constant radius of curvature \cite{Chen:2009zp}, the form of the transfer vertex $\rho \dot{\varphi} \sigma $ is generic~\cite{Assassi2013} in the sense that it comes from the leading order allowed interaction between a (nearly) shift symmetric inflaton and a (spectator) second scalar. Quasi-single field inflation non-Gaussianity was studied in detail with a cubic self-interaction for $\sigma$ in \cite{Chen2010a,Chen:2009zp}. The relevant diagrams for computing the contribution of the spectator field to the observed correlation functions for that case are diagrammatically depicted in Fig.~\ref{fig:QSFI-interaction-terms}. The transfer vertex will allow $\sigma$ to contribute to the power spectrum as well as to generate a three-point function for $\varphi$ from the $\sigma^3$ interaction.

Since the self-interactions of the inflaton mode will have no significant long-short coupling, in this paper we are only interested in computing correlations coming from self-interaction vertices of $\sigma$. The spectator scalar may well have additional self-interactions beyond the cubic term. Here we will consider
\begin{equation}
\label{eq:V_sigma}
	V(\sigma) = \mu \sigma^3+\frac{\lambda}{4!}\sigma^4 + \frac{g}{5!}\sigma^5+\frac{h}{6!}\sigma^6 + \dots.
\end{equation}
In the next section we compute the correlations generated by these interactions in the limits appropriate for determining the effects of long-short mode coupling on $n$-point functions in sub-volumes.

\subsection{Quantization and in-in formalism}
To derive the late time correlation functions from the inflationary scenario above we use the in-in formalism ~\cite{Weinberg2005,Maldacena:2002vr}, where one computes expectation values of fields at a given time. For example, if $Q(\tau)$ is a product of field operators whose correlation function we want to evaluate at conformal time $\tau$, we compute
\begin{equation}
	\langle Q(\tau) \rangle = \left\langle 0 \right| \left[ \overline{T} \exp \left( i \int_{-\infty}^{\tau} d\tau^{\prime} H_{I}(\tau^{\prime}) \right) \right] Q(\tau) \left[ T \exp \left( -i \int_{-\infty}^{\tau} d\tau^{\prime} H_{I}(\tau^{\prime}) \right) \right] \left| 0 \right\rangle,
\label{eq:in-in-prescription}
\end{equation}
where $H_I (\tau)$ is the interaction Hamiltonian in the interaction picture, $T, \overline{T}$ are the time ordering and anti-time ordering operators. Similarly to the procedure for computing scattering amplitudes, the expression in Eq.(\ref{eq:in-in-prescription}) can be evaluated in perturbation theory by expanding the exponentials, and gathering same-order terms together. One way of arranging those terms together, convenient for our purposes here, is the commutator-form:
\begin{equation}
	\langle Q(\tau) \rangle = \sum_{n} i^n \int_{-\infty}^{\tau} d\tau_1 \int_{-\infty}^{\tau_1} d\tau_2 \dots \int_{-\infty}^{\tau_{n-1}} d\tau_n \left\langle 0 \right| [H_I(\tau_n),[H_I(\tau_{n-1}), \dots ,[H_I(\tau_1), Q(\tau)]\dots ]] \left| 0 \right\rangle.
\label{eq:in-in-commutator}
\end{equation}
The standard $i\epsilon$ prescription should be applied to the lower limits of the time integrals to project onto the Bunch Davies vacuum at early time.

The perturbation fields are quantized in the usual way, with Fourier components in the interaction picture given by:
\begin{align}
\label{eq:aadagger}
	\varphi_{\vec{k}} (\tau) = u_{k}(\tau) a_{\vec{k}} + u^{*}_{k} (\tau) a^{\dagger}_{\vec{k}}, \\\nonumber
	\sigma_{\vec{k}} (\tau) = v_{k}(\tau) b_{\vec{k}} + v^{*}_{k} (\tau) b^{\dagger}_{\vec{k}} \;.
\end{align}
Here $a_{\vec{k}}$ and $b_{\vec{k}}$ obey the usual commutation relations. Assuming a quasi-de Sitter background, the mode function for $\sigma$ satisfying the equations of motion of the quadratic Lagrangian Eq.(\ref{eq:Lagrangian-QSFI}) (with $\rho\ll H$) is
\begin{equation}
{\rm spectator\; scalar:}\;\;\;\;	v_k (\tau ) = \frac{H \sqrt{\pi}}{2 \sqrt{k^3}} (-k\tau)^{3/2} H_{\nu}^{(1)} (-k\tau)\;.
\label{eq:mode-fct-v}
\end{equation}
Note that the order of the Hankel function of the first kind, $ H_{\nu}^{(1)}$, is determined by the parameter $\nu$, which depends on the mass of the spectator field. In the limit of scale-invariant background evolution, this is:
\begin{equation}
	\nu \equiv \sqrt{\frac{9}{4} - \frac{m^2}{H^2}}\;.
\label{eq:def-nu}
\end{equation}
The mode function for the inflaton fluctuations~$\varphi$ are:
\begin{equation}
{\rm inflaton:}\;\;\;\;u_{k} (\tau) = \frac{H}{\sqrt{2k^3}} (1+ik\tau) e^{-ik \tau} \;.
\label{eq:mode-fct-u}
\end{equation}
and can be obtained -- up to a phase factor -- from the mode functions for $\sigma$ by taking the massless limit ($\nu \to 3/2$).
We will frequently need the late time approximation for the mode functions, appropriate for when the modes are well outside the Hubble radius. The inflaton mode function in this limit is just $H/\sqrt{2k^3}$, while for the spectator scalar it is
\begin{equation}
	\lim_{\mid k \tau \mid \to 0} v_k (\tau) \sim -i \frac{2^{\nu} \Gamma(\nu)}{2 \sqrt{\pi}} \frac{H}{\sqrt{k^3}} (-k\tau)^{3/2-\nu}.
	\label{eq:mode-fct-v-approx}
\end{equation}

\subsection{The \texorpdfstring{$N$}{N}-point functions from self-interactions of the hidden sector field}
\label{sec:kernels}
In this section we apply the formalism above to compute the correlation functions of the adiabatic mode at late times. That is, we use the $n$-field contact interactions in the interaction Hamiltonian to evaluate $\langle Q_I\rangle=\langle\zeta^n\rangle$ for $n\geq 3$ when all the modes are well outside the inflationary Hubble radius. 

We will assume homogeneous and isotropic fluctuations, so that the $n$-point functions depend on $n-1$ independent momenta. For example, the power spectrum $P(k)$, bispectrum $B(\kv_1,\kv_2,\kv_3)$ and trispectrum $T(\kv_1,\kv_2,\kv_3,\kv_4)$ are defined from the two-, three- and four-point functions, respectively, in the following way:
\begin{align}
	\label{eq:biandtri}
	\langle\zeta(\mathbf{k}_1)\zeta(\mathbf{k}_2) \rangle=& (2\pi)^3 \, \delta^{(3)}(\mathbf{k}_1+\mathbf{k}_2) \, P(k_1),\\
	\langle \zeta(\kv_1) \zeta(\kv_2) \zeta( \kv_3)\rangle=& (2\pi)^3 \, \delta^{(3)}(\mathbf{k}_1+\mathbf{k}_2+\mathbf{k}_3)  \, B(\kv_1,\kv_2,\kv_3), \nn\\
	\langle \zeta(\mathbf{k}_1)\zeta(\mathbf{k}_2)\zeta(\mathbf{k}_3) \zeta(\mathbf{k}_4)\rangle= & (2\pi)^3 \, \delta^{(3)}(\mathbf{k}_1+\mathbf{k}_2+\mathbf{k}_3+\mathbf{k}_4) \, T(\kv_1,\kv_2,\kv_3,\kv_4)\nn
\end{align}
For higher order correlations we define
\begin{equation}
	\langle \zeta(\mathbf{k}_1)\zeta(\mathbf{k}_2)...\zeta(\mathbf{k}_n)\rangle\equiv (2\pi)^3 \, \delta^{(3)}(\sum_{i=1}^{n}\mathbf{k}_i)  \, F_n(\kv_1,\kv_2,...\kv_n).
\end{equation}

Evaluating the exact higher order correlation functions in quasi-single field with additional interaction terms requires lengthy calculations because of vertices transferring power between $\sigma$ and $\varphi$ ($n+1$ integrals and commutators in Eq.(\ref{eq:in-in-commutator})). However, for our purposes only the degree of long-short coupling in soft limits is required, which simplifies the calculation. In this section we review the previously calculated quasi-single field bispectrum and then use simple arguments (well understood for the bispectrum in e.g., \cite{Baumann2011b}) to determine the relevant features of higher order correlation functions. Additional details justifying these arguments for the trispectrum can be found in Appendix \ref{app:trisp-estimates}.

\subsubsection{Bispectrum}
\label{sec:kernels-bisp}
The full bispectrum from the cubic interaction only, $V(\sigma) = \mu \sigma^3$, was calculated in \cite{Chen2010a}, who also gave an approximate form that accurately captures the behavior in the squeezed limit:
\begin{equation}
B_{\zeta}(k_1,k_2,k_3) =  f_{NL} \frac{6 \cdot 3^{3/2}}{N_\nu(\alpha/27)} \frac{(2 \pi^2 \Delta_{\zeta}^2)^2}{(k_1k_2k_3)^{3/2}(k_1 + k_2 + k_3)^{3/2}} N_\nu\left( \frac{\alpha k_1 k_2 k_3}{(k_1 + k_2 + k_3)^3} \right)
\label{eq:ansatz-bispectrum}
\end{equation}
where a scale-invariant power spectrum has been assumed to define $\Delta_\zeta^2 \equiv k^3 P_\zeta(k)/2\pi^2$. The function $N_\nu$ is a Bessel function of the second kind (also known as a Neumann function) and numerically fitting the shape ansatz above to the exact result determined the numerical parameter $\alpha \simeq 8$~\cite{Chen2010a}. In terms of the parameters in the Lagrangian, $f_{NL}$ is
\begin{equation}
	f_{NL}=-\frac{9}{20}f(\nu) \Delta_\zeta^{-1} \left( \frac{\rho}{H} \right)^3 \left(\frac{\mu}{H} \right),
	\label{eq:def_fnl}
\end{equation}
chosen to match the normalization of the local ansatz in the $k_1=k_2=k_3$ configuration. Note that we have departed from the original definitions by moving a factor of $\frac{3}{5}$ into $\fnl$ since it will simplify our convention for generic non-Gaussian fields in Eq.(\ref{eq:zng-expansion-in-Z}).
The function $f(\nu)$ is positive and monotonic (order 1 for $\nu\lesssim0.9$ and grows rapidly to $\mathcal{O}(100)$ as $\nu\rightarrow\frac{3}{2}$). It is plotted in \cite{Chen2010a}.\footnote{There it is called $\alpha(\nu)$.} The factor $\mu/H$ is due to the hidden sector cubic coupling, while the factors of $\rho/H$ come from the three transfer vertices that convert the perturbations in the hidden field back to perturbations of the inflaton. Note that although $\frac{\rho}{H},\;\frac{\mu}{H}\ll1$, the factor of $\Delta^{-1}_{\zeta}$ is large and $f_{\text{NL}}>1$ is in the allowed parameter space.

Our interest here is in the so-called squeezed limit of the bispectrum, where one of the momenta is much smaller than the other two. Denoting the long and short wavelength modes $k_L=k_1\ll k_2\approx k_3\equiv k_S$ and expanding Eq.(\ref{eq:ansatz-bispectrum}) in this limit gives
\begin{equation}
	\lim_{k_1\ll k_2\approx k_3} B(k_1,k_2,k_3) \propto P(k_S) P(k_L) \left( \frac{k_L}{k_S} \right)^{3/2-\nu}\;,
\label{eq:bisp-in-squeezed-limit}
\end{equation}
where we have used $N_{\nu}(x)\overset{{x\rightarrow0}}{\longrightarrow}-\frac{\Gamma(\nu)}{\pi}(\frac{2}{x})^{\nu}$. Notice that when the fluctuation of the hidden sector field is massless ($\nu=3/2$), the above expression recovers the local type bispectrum. When $\sigma$ is massive ($\nu<\frac{3}{2}$) this bispectrum is less divergent as $k_L\rightarrow0$ and so has a weaker long-short mode coupling than the local ansatz. If a non-zero bispectrum of the quasi-single field type is detected, measuring its scaling with $k_L$ in the squeezed limit would amount to measuring the mass of the hidden sector fluctuation (although note that even with a detection, the observational precision on this number is likely to be quite poor for the near future).

\subsubsection{Trispectrum}
\label{sec:kernels-trisp}
Adding the interaction term proportional to $\sigma^4$ introduces a new trispectrum to the original quasi-single field scenario. Rather than performing a complete in-in derivation of the exact trispectrum, we will use dimensional analysis in a few simple limits to derive the amplitude and momentum dependence of the trispectrum in squeezed configurations (analogous to the arguments presented in \cite{Baumann2011b} for the bispectrum). From these estimations we derive an ansatz for the trispectrum, convenient for our calculations. The amplitude of the trispectrum was previously discussed in \cite{Chen:2009zp}. Some details of the in-in result can be found in Appendix~\ref{app:trisp-estimates}.

\begin{figure}
	\centering
	\begin{tikzpicture}
		\draw[thick, dashed] (0,0) -- (-.5,0.5);
		\draw[thick, dashed] (0,0) -- (-.5,-0.5);
		\draw[thick, dashed] (0,0) -- (.5,0.5);
		\draw[thick, dashed] (0,0) -- (.5,-0.5);
		\draw[thick] (0.5,0.5) -- (1,1);
		\draw[thick] (0.5,-0.5) -- (1,-1);
		\draw[thick] (-0.5,0.5) -- (-1,1);
		\draw[thick] (-0.5,-0.5) -- (-1,-1);
		\draw (-1.2,1.2) node{$k_1$};
		\draw (-1.2,-1.2) node{$k_2$};
		\draw (1.2,1.2) node{$k_3$};
		\draw (1.2,-1.2) node{$k_4$};
	\end{tikzpicture}
\caption{The four-point interaction vertex. As before solid lines represent the $\varphi$ field while dashed lines are $\sigma$.}
\label{fig:int-trisp-estimate}
\end{figure}
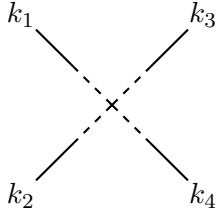

The four-point correlation function from the quartic interaction vertex is depicted in \fref{fig:int-trisp-estimate} and the late time ($\tau=0$) result can schematically be expressed as:
\begin{equation}
\begin{split}
    \langle \varphi_{\mathbf{k}_1} \varphi_{\mathbf{k}_2} \varphi_{\mathbf{k}_3} \varphi_{\mathbf{k}_4} \rangle\bigr|_{\tau=0}  =  &(2 \pi)^3 \delta^3(\kv_1+\kv_2+\kv_3+\kv_4)\left(\varphi_{\mathbf{k}_1} \varphi_{\mathbf{k}_2} \varphi_{\mathbf{k}_3} \varphi_{\mathbf{k}_4} \right)\bigr|_{\tau=0} \\
    &\times \left( \displaystyle \prod_{i=1}^{4} \int d\tau_i a^3 \rho \, \varphi^{\prime}_{k_i} \sigma_{k_i} \right) \left( \int d\tau a^4 \lambda \, \sigma_{k_1} \sigma_{k_2} \sigma_{k_3} \sigma_{k_4} \right).
\end{split}
\label{eq:int-squeezed-four-points}
\end{equation}
Notice that the integrals associated with the mixing term (inside the first set of parenthesis on the second line above) each depend only on a single momenta and are dimensionless (once fields are written in terms of mode functions, Eq.(\ref{eq:aadagger})). So, these integrals should contribute no ratios of momenta and can be approximated by $\rho / H$. The momentum dependence of the trispectrum can be extracted from the remaining terms. The multiplicative factor on the first line of Eq.~\eqref{eq:int-squeezed-four-points} is just related to the power spectrum of $\varphi$:
\begin{equation}
	\left( \varphi_{\mathbf{k}_1} \varphi_{\mathbf{k}_2} \varphi_{\mathbf{k}_3} \varphi_{\mathbf{k}_4}\right)\bigr|_{\tau=0}  \sim \frac{\Delta^{4}_{\varphi}}{(k_1 k_2 k_3 k_4)^{3/2}}.
\end{equation}

The remaining integral over the $\sigma$ self-interaction, the last parenthesis of Eq.(\ref{eq:int-squeezed-four-points}), depends on all momenta. In general, it is clear from the oscillatory form of the mode functions for $|k\tau|\gg1$ that contributions from modes deep in the UV will be suppressed. So, in momentum configurations where there is a largest momenta, the dominant contribution to the integral will come when that mode finally crosses the horizon, $\tau\sim -1/k_{UV}$. Here we are primarily interested in the bispectrum induced in sub-volumes when one mode of the trispectrum is unobservable (e.g., $k_1$ is super-Hubble). Furthermore, we are interested in the squeezed limit of that bispectrum so the momentum configuration is $k_1\ll k_2\ll k_3\approx k_4$. (This momentum configuration can also be used to work out the correction to the power spectrum in biased sub-volumes when both $k_1$ and $k_2$ are super-Hubble.) In that case the integral above is dominated by $\tau\approx -1/k_3=-1/k_4\equiv k_S$, and the result is
\begin{align}
\label{eq:eval_int_integral}
	\lim_{k_1, k_2 \ll k_3\approx k_4} \int d \tau a^4 \lambda \, \sigma_{k_1} \sigma_{k_2} \sigma_{k_3} \sigma_{k_4} & \sim \int d \tau \lambda (- \tau )^{2-4\nu} (k_1 k_2 k_3 k_4)^{-\nu} \nonumber\\
	& \sim \lambda \, \ k_{L_1}^{-\nu} k_{L_2}^{-\nu} (k_S)^{2\nu-3},
\end{align}
where in the last line we have labelled the two (not necessarily equal) long modes $k_{L_1}$, $k_{L_2}$ and the short mode $k_S$.

Putting all the pieces together, and converting $\varphi$ to $\zeta$, gives the trispectrum in the $k_1,\; k_2\ll k_3\approx k_4$ configuration
\begin{equation}
	\lim_{k_1, k_2 \ll k_3\approx k_4} T(k_1,k_2,k_3,k_4) \propto \frac{\Delta_{\zeta}^6}{(k_1 k_2 k_3)^3} \ (\lambda \Delta^{-2}_{\zeta}) \left(\frac{\rho}{H} \right)^4  \left( \frac{k_1}{k_3} \right)^{3/2-\nu} \left( \frac{k_2}{k_3} \right)^{3/2-\nu}.
\label{eq:trisp-dsqueezed}
\end{equation}
By analogy with the ansatz for the bispectrum, we can guess an approximate form for the trispectrum that is useful because it is symmetric in all momenta:
\begin{equation}
T(k_1,k_2,k_3,k_4)\propto\frac{g_{NL} \, \Delta_{\zeta}^6}{(k_1k_2k_3k_4)^{3/2}(k_1 + k_2 + k_3+k_4)^{3}} N_\nu\left( \frac{\alpha_4 k_1 k_2 k_3k_4}{(k_1 + k_2 + k_3+k_4)^4} \right)\;,
\label{eq:trispectrum-ansatz}
\end{equation}
where $\alpha_4$ is some numerical constant and $\gnl\propto\lambda(\rho/H)^4\Delta_{\zeta}^{-2}$. 

This ansatz also has the correct scaling in the $k_1,\; k_2\ll k_3\approx k_4$ case, as well as when $k_1\ll k_2\approx k_3\approx k_4$. When one momenta is very soft and the others are in an equilateral configuration the integral in Eq.(\ref{eq:eval_int_integral}) can be re-evaluated and the final result agrees with expanding Eq.(\ref{eq:trispectrum-ansatz}):
\begin{equation}
	\lim_{k_1\ll k_2\approx k_3\approx k_4} T(k_1,k_2,k_3,k_4)\propto \frac{\Delta_{\zeta}^6}{(k_1 k_2 k_3)^3} \ \lambda \left(\frac{\rho}{H} \right)^4 \Delta^{-2}_{\zeta}  \left( \frac{k_1}{k_2} \right)^{3/2-\nu}.
\label{eq:trisp-ssqueezed}
\end{equation}
This limit, together with the previous one, shows that when $\nu=3/2$ the trispectrum from the $\sigma^4$ interaction has the same limits as the usual local type $\gnl$ trispectrum. The numerical coefficients in Eq.(\ref{eq:trispectrum-ansatz}) could be chosen to match the normalization to that shape. (We have not checked how well the ansatz in Eq.(\ref{eq:trispectrum-ansatz}) works in more general configurations.)

There is also a trispectrum from the cubic interaction alone, coming from a diagram with two cubic interactions connected by a $\sigma$ line. There will be many such exchange diagrams at higher orders. These are very interesting, but we wish to focus here on the effects of the contact terms from the series of interaction terms in Eq.(\ref{eq:V_sigma}), which in some ways mimics the series expansion of the local ansatz. Appendix \ref{app:exchange} discusses the 4-point exchange diagram from the cubic interaction in more detail, and we will comment later on incorporating it in the expression for the full non-Gaussian field, but otherwise leave a complete discussion of these diagrams for future work.

\subsubsection{Higher order correlation functions}
In the previous two subsections~\ref{sec:kernels-bisp} and~\ref{sec:kernels-trisp} we studied various squeezed limit(s) of the bispectrum and the trispectrum. Here, we will generalize those results to arbitrarily high $n-$point functions generated by self-interaction terms in the hidden sector perturbation field. This will be needed in order to study the contribution from higher order correlation functions on lower order ones through mode-mode coupling.

In particular, we consider the $n$-point function arising from a contact diagram with an interaction term
\begin{equation}
	V_n (\sigma) = \frac{\lambda_n}{n!} \sigma^n,
\end{equation}
and $n$ transfer vertices. This can easily be estimated using similar arguments as in the previous subsections. The schematic expression of the $n$-point function is given by:
\begin{equation}
	\langle \zeta_{\mathbf{k_1}} \ldots \zeta_{\mathbf{k_n}} \rangle \sim \delta^3\left(\sum_i^{n}\kv_i\right)\;\zeta_k^n\bigr|_{\tau=0}  \; \prod_{i=1}^{n} \left(\int d\tau_i a^3 \rho \varphi^{\prime} \sigma \right) \; \times \int d\tau a^4 \lambda_n \sigma^n \; .
\end{equation}
Again the product of all the integrals associated with the transfer vertex can be approximated by $(\rho/H)^n$. If we consider a configuration with $\ell$ long modes $k_{L_i}$ (which can be either super-Hubble or long modes within a sub-volume) and $n-\ell$ short modes assumed to be all of equal length $k_S$, the pre-factor can be written as
\begin{equation}
	\zeta_k^n\bigr|_{\tau=0}  \propto \frac{\Delta_{\zeta}^{n}}{k_S^{3(n-\ell)/2}\prod_{i=1}^{\ell}k_{L_i}^{3/2}} \;.
\end{equation}
The remaining integral will be suppressed except when the most UV mode is nearly at the Hubble scale, $-\tau\approx1/k_S$:
\begin{align}
	\int d\tau a^4 \lambda_n \sigma^n & \sim \int d\tau \frac{\lambda_n}{(-H\tau)^4} H^n \frac{(-\tau)^{(3/2-\nu)n}}{\prod_{i=1}^{\ell}k_{L_i}^{\nu}\,k_S^{(n-\ell)\nu}} \nonumber \\
	& \sim \lambda_n H^{n-4} \prod_{i=1}^{\ell}k_{L_i}^{-\nu} \; \int d\tau  k_S^{-(n-\ell)\nu} (-\tau)^{(3/2-\nu)n - 4} \nonumber \\
	& \sim \lambda_n H^{n-4}  \prod_{i=1}^{\ell}k_{L_i}^{-\nu} \; k_S^{\ell \nu -3/2 n +3}.
\end{align}
Putting everything together gives the general expression for the $n$-point correlation function with $\ell$ modes taken to be long:
\begin{align}
	\lim_{k_1,\dots k_{\ell} \ll k_{\ell+1}\approx\dots\approx k_4}F_n(\kv_1,\dots,\kv_n) & \propto \lambda_n H^{n-4} \Delta_{\zeta}^{n} \left( \frac{ \rho}{H}\right)^n k_S^{-3(n-\ell-1)} \prod_{i=1}^{\ell}k_{L_i}^{-3}\left( \frac{k_{L_i}}{k_S} \right)^{(3/2-\nu)} \nonumber \\
	& \sim  \lambda_n H^{n-4} \Delta_{\zeta}^{2-n}\left( \frac{ \rho}{H}\right)^nP^{n-\ell-1}(k_S) \prod_{i=1}^{\ell}P(k_{L_i})\left( \frac{k_{L_i}}{k_S} \right)^{(3/2-\nu)}.
	\label{eq:n-point-correlation}
\end{align}
This behavior can be captured by a template similar to the lower order expressions
\begin{equation}
\label{eq:npoint}
F_n(\kv_1,\dots,\kv_n) \propto  \left( k_1 \, k_2 \, \ldots k_n \right)^{-\frac{3}{2}} \left(k_1 + k_2 + \ldots + k_n \right)^{3-\frac{3}{2} n} N_\nu \left(\frac{\alpha_n k_1 \, k_2 \, \ldots k_n  }{\left(k_1 + k_2 + \ldots + k_n \right)^n} \right)
\end{equation}
where $\alpha_n$ is a numerical coefficient that can be chosen to help fit the exact result.

\section{Cosmic variance from super-horizon modes}
\label{sec:shiftBS}
Once the post-inflationary correlation functions have been determined, it is a purely mathematical exercise to compute the statistics in spatial sub-volumes at a fixed time.  For this purpose, we introduce in Section~\ref{sec:late-time-correlation-functions} a formalism to build up the non-Gaussian field from its correlation functions. We introduce a split between long and short modes to derive the non-Gaussian field observed in sub-volumes. Of course, since in the quasi-single field case we also know the dynamical model generating the fluctuations, we could just as well do the whole calculation within the in-in formalism. We do an example in-in calculation in Section \ref{sec:inin_split} to confirm that the two methods agree. In addition, Section \ref{sec:one-loop} in the Appendix contains an in-in calculation that demonstrates aspects of the dynamical calculation that are distinct from the purely statistical effects of sub-sampling.

\subsection{Late-time correlation functions and superhorizon modes}
\label{sec:late-time-correlation-functions}

In order to provide a framework for our calculations, we first establish our notation for generic non-Gaussian fields in the post-inflationary universe. If the correlation functions of the scalar metric fluctuation $\zeta$ are specified on a spatial slice at some early time (but after reheating and any other era that could have transferred isocurvature modes into the adiabatic mode) it is straightforward to determine the distribution of correlation functions observed in sub-volumes.

The non-Gaussian mode can be expressed as a sum of terms $Z_n$ that are local or non-local functionals of $n$ Gaussian random fields $\zeta_G$:
\begin{equation}
\label{eq:zetaNGx}
	\zeta_{\text{NG}} (\mathbf{x}) = Z_1[ \zeta_{\text{G}} (\mathbf{x})] + f_{\text{NL}} Z_2[\zeta_{\text{G}}(\mathbf{x})]+ g_{\text{NL}} Z_3[\zeta_{\text{G}}(\mathbf{x})]+ \dots
\end{equation}
In Fourier space, this series is
\begin{equation}
	\zng(\mathbf{k})=Z_1(\mathbf{k})+ \fnl Z_2(\mathbf{k})+ \gnl Z_3(\mathbf{k})+\dots
\label{eq:zng-expansion-in-Z}
\end{equation}
where $Z_1(\mathbf{k})=f_L(\mathbf{k}) \zg(\mathbf{k})$ is just proportional to the Gaussian field\footnote{ In the absence of any mode-coupling effects the coefficient of the linear term, $f_L$, can just be absorbed into the variance of the Gaussian field $\zeta_G$. However, in what follows we would like Eq.(\ref{eq:zng-expansion-in-Z}) to apply in cases where the amplitude of fluctuations can differ in sub-volumes. The notation for that case is clearer if we allow for the possibility $f_L\neq1$ and $f_L$ momentum-dependent.}. The higher order terms are convolutions of $n$ Gaussian fields. For example,
\begin{align}
\begin{split}
	Z_2(\mathbf{k}) & = \frac{1}{2!(2\pi)^3}\int d^3p_1 d^3p_2 \,[\zg(\mathbf{p}_1)\zg(\mathbf{p}_2)-\langle\zg(\mathbf{p}_1)\zg(\mathbf{p}_2)\rangle] \\
	 & \hspace{2cm} \times N_2(\mathbf{p}_1, \mathbf{p}_2,\mathbf{k})\,\delta^{(3)}(\mathbf{k}-\mathbf{p}_1-\mathbf{p}_2) \label{eq:def-Z_2}
\end{split} \\
\begin{split}
	Z_3(\mathbf{k}) & = \frac{1}{3!(2 \pi)^6}\prod_{\ell=1}^3\int  d^3 p_\ell \left[\zg(\mathbf{p}_1)\zg(\mathbf{p}_2)\zg(\mathbf{p}_3)-\sum_{\substack{i=1\\k\neq j\neq i}}^3\zg(\mathbf{p}_i)\langle\zg(\mathbf{p}_j)\zg(\mathbf{p}_k)\rangle\right]\\
	&\hspace{2cm}\times N_3(\mathbf{p}_1,\mathbf{p}_2,\mathbf{p}_3,\mathbf{k})\,\delta^{(3)}(\mathbf{k}-\sum_{\ell=1}^3\mathbf{p_{\ell})}\;. \label{eq:def-Z_3}
\end{split}
\end{align}
The kernels $N_n$ are symmetric in the first $n$ momenta and are chosen to reproduce the tree level $(n+1)$-point function. The structure of the subtracted expectation values ensures that $\zng$ has mean zero and that the non-linear terms only contribute to the connected parts of the correlations. The coefficients $\fnl$ and $\gnl$ are numbers which can only be unambiguously defined when the kernels are scale-invariant. In that case, the kernels can be normalized so that $\fnl$ and $\gnl$ agree with, eg, the usual coefficients of the local templates\footnote{This works for kernels with non-vanishing equilateral limits, which is true of those we consider in this paper.} (although notice that to keep the notation uncluttered we have not separated out the usual factors of $\frac{3}{5}$ used since $\fnl$, etc. are most often defined in the matter era Bardeen potential).

The effective non-Gaussian field that gives the statistics observed in a \emph{sub-volume} can be found by considering Eq.\eqref{eq:zetaNGx} restricted to a spatial region of linear size $\sim L$. This field is approximately the same as that obtained from the simpler procedure of considering Eq.\eqref{eq:zng-expansion-in-Z} with some modes having momenta smaller than a cut-off $k_0\approx 2\pi/L$. We define $Z_i^{\rm \ell\,long}(\kv)$ as $Z_i$ with $\ell$ long wave-length modes. For example,
\begin{align}
\begin{split}
Z_2^{\rm 1\,long}(\kv) & = \frac{1}{2!(2\pi)^3}\int_{p_1>k_0} d^3p_1\int_{p_2<k_0} d^3p_2 \,[\zg(\mathbf{p}_1)\zg(\mathbf{p}_2)-\langle\zg(\mathbf{p}_1)\zg(\mathbf{p}_2)\rangle] \\
 & \hspace{2cm} \times N_2(\mathbf{p}_1, \mathbf{p}_2,\mathbf{k})\,\delta^{(3)}(\mathbf{k}-\mathbf{p}_1-\mathbf{p}_2)\;,
 \end{split} \\
\begin{split}
Z_3^{\rm 2\,long}(\kv)&=\frac{1}{3!(2 \pi)^6}\,\int_{p_1>k_0} d^3p_1\;\prod_{\ell=2}^3\int_{p_i<k_0} d^3 p_\ell \,\\
&\hspace{2cm}\times\left[\zg(\mathbf{p}_1)\zg(\mathbf{p}_2)\zg(\mathbf{p}_3)-\sum_{\substack{i=1\\k\neq j\neq i}}^3\zg(\mathbf{p}_i)\langle\zg(\mathbf{p}_j)\zg(\mathbf{p}_k)\rangle\right]\\
&\hspace{2cm}\times N_3(\mathbf{p}_1,\mathbf{p}_2,\mathbf{p}_3,\mathbf{k})\,\delta^{(3)}(\mathbf{k}-\sum_{\ell=1}^3\mathbf{p_{\ell})}\;.
\end{split}
\end{align}
Then the observed field will be
\begin{equation}
\begin{split}
	\zeta^{\rm obs}_{\rm NG}(\kv)&=\zeta_G(\kv)\left[1+2\fnl Z_2^{\rm 1\,long}(\kv)+3\gnl Z_3^{\rm 2\,long}(\kv)+4h_{\rm NL}Z_4^{\rm 3\,long}(\kv)+\dots\right]\\
	&\quad +\left[\fnl Z_2^{\rm 0\,long}(\kv)+3\gnl Z_3^{\rm 1\,long}(\kv)+6h_{\rm NL}Z_4^{\rm 2\,long}(\kv)+\dots\right]\\
	&\quad +\left[\gnl Z_3^{\rm 0\,long}(\kv) +4h_{\rm NL} Z_4^{\rm 1\,long}(\kv)\right]+\dots
\end{split}
\end{equation}
The numerical pre-factors account for the fact that the integrals in the $Z_n$ are symmetric in the $p_i$, so that equivalent contributions come from choosing any of the momenta (not just the last $\ell$ of the $p_i$) to be the long-wavelength modes. 

The first line in the equation above is the linear field observed in the sub-volume, so to compute the observed non-Gaussian correlations in terms of the observed power spectrum, the $Z_i$ should be re-expressed in terms of this field. That shift can be absorbed into a re-definition of the kernels $N_i$. In other words, an observer in the sub-volume sees statistics generated by
\begin{equation}
\zng^{\text{obs}}(\mathbf{k})=\chi_G(\kv)+ \fnl^{\rm obs} Z^{\rm obs}_2[\chi_G(\kv)]+ \gnl^{\rm obs} Z^{\rm obs}_3[\chi_G(\kv)]+\dots
\label{eq:expansion-zeta-in-chis}
\end{equation}
where
\begin{equation}
\chi_G(\kv)\equiv \,f_L(\kv)\zeta_G(\kv)=\left[1+2\fnl Z_2^{\rm 1\,long}(\kv)+3\gnl Z_3^{\rm 2\,long}(\kv)+\dots\right]\zeta_G(\kv).
\label{eq:expansion-chi-in-Z}
\end{equation}
The functional $Z_2^{\text{obs}}$ is defined by
\begin{equation}
\begin{split}
	\fnl^{\rm obs}Z_2^{\rm obs}(\kv)=&\frac{f_{\text{NL}}}{2!(2\pi)^3}\int_{p_1>k_0} d^3p_1\int_{p_2>k_0} d^3p_2 \,[\chi_G(\mathbf{p}_1)\chi_G(\mathbf{p}_2)-\langle\chi_G(\mathbf{p}_1)\chi_G(\mathbf{p}_2)\rangle]\\
	& \hspace{2cm} \times N^{\rm obs}_2(\mathbf{p}_1, \mathbf{p}_2,\mathbf{k})\,\delta^{(3)}(\mathbf{k}-\mathbf{p}_1-\mathbf{p}_2)
\end{split}
\label{eq:Z2-obs-def}
\end{equation}
with an effective kernel depending on the higher order functionals containing the proper number of long modes
\begin{align}
\label{eq:N2-obs}
	N^{\rm obs}_2(\mathbf{p}_1, \mathbf{p}_2,\mathbf{k})=&\frac{N_2(\mathbf{p}_1, \mathbf{p}_2,\mathbf{k})}{f_L(\mathbf{p}_1)f_L(\mathbf{p}_2)}\\
	&+ \sum_{n=3} \frac{n!}{2!(n-2)!} \frac{c^{(n)}_{\text{NL}}}{f_{\text{NL}}} \left( \prod_{i=3}^{n} \int_{p_i < k_0} \frac{d^3 p_i}{(2\pi)^3 } \; \frac{\chi_G(\mathbf{p}_i)}{f_L(\mathbf{p}_i)} \right) \frac{N_n(\mathbf{p}_1,\dots,\mathbf{p}_n,\mathbf{k})}{f_L(\mathbf{p}_1) f_L(\mathbf{p}_2)}\nn
\end{align}
where we have denoted $\gnl=c^{(3)}_{\text{NL}}$, $h_{\rm NL}=c^{(4)}_{\text{NL}}$, etc. Higher order $Z_i^{\text{obs}}$ are defined similarly:
\begin{align}
\label{eq:Nn-obs}
	N^{\rm obs}_n(\mathbf{p}_1, \dots, \mathbf{p}_n,\mathbf{k})=&\frac{N_n(\mathbf{p}_1, \dots,\mathbf{p}_n,\mathbf{k})}{f_L(\mathbf{p}_1) \dots f_L(\mathbf{p}_n)}\\
	&+ \sum_{\ell=n+1} \frac{\ell! }{n!(\ell-n)!} \frac{c^{(\ell)}_{\text{NL}}}{c^{(n)}_{\text{NL}}}\left( \prod_{i=n+1}^{\ell} \int_{p_i < k_0} \frac{d^3 p_i}{(2\pi)^3} \; \frac{\chi_G(\mathbf{p}_i)}{f_L(\mathbf{p}_i)} \right) \frac{N_\ell(\mathbf{p}_1,\dots,\mathbf{p}_\ell,\mathbf{k})}{f_L(\mathbf{p}_1) \dots f_L(\mathbf{p}_n)}.\nn
\end{align}

In the next subsections we work out the kernels $N_2$, $N_3$ and higher order ones when the inflaton has the quasi-single field coupling to an additional light scalar with arbitrary non-derivative self-interactions. Because our goal is to understand how long wavelength fluctuations affect the power spectrum and squeezed limit of the bispectrum in biased sub-volumes, we will not need the exact expressions for every momentum configuration. One should keep in mind that the results below are not valid away from the squeezed limits.

\subsection{Variance of scalar spectral index and amplitude of power spectrum}
\label{sec:pheno-B-on-P}
We can now use the late time correlation functions computed in Section \ref{sec:QSF} to express the non-Gaussian perturbation $\zeta_{\text{NG}}$ as an expansion in terms of a Gaussian field, Eq.(\ref{eq:zng-expansion-in-Z}).

Since we begin by considering the field in the entire inflationary volume, $Z_1(\kv)=\zeta_G(\kv)$. Each higher order $Z_n(\mathbf{k})$ involves integrals over momentum, a delta-function for momentum conservation, and a kernel. We choose the two first kernels $N_2$ and $N_3$ to reproduce the squeezed limits of the bispectrum and trispectrum derived in the previous section\footnote{The non-Gaussian field built with these kernels is still only approximately that of the full quasi-single field model. The quadratic kernel will generate a contribution to the trispectrum that is not present in the model, which we will ignore because it is suppressed by two factors of $\rho/H$ (and could be explicitly canceled by adding an appropriate piece to $N_3$). In addition, quasi-single field contains contributions to the correlations from exchange diagrams such as the trispectrum piece discussed in Appendix \ref{app:exchange}. This ansatz for $N_3$ captures part, but not all of that diagram. Similarly, our higher order kernels will not capture all the contributions from all exchange diagrams. However, here we want to focus on the effects of the contact diagrams from new interactions so we leave the full discussion of exchange diagrams for future work.}:
\begin{align}
\label{eq:N23}
	N_2(p_1,p_2,k) & \propto \frac{(p_1+p_2+k)^{3\nu -3/2}}{(p_1p_2k)^{3/2+\nu}} p_1^3 p_2 ^3, \\
	N_3(p_1,p_2,p_3,k) & \propto \frac{(p_1+p_2+p_3+k)^{4\nu - 3}}{(p_1p_2p_3k)^{3/2+\nu}} p_1^3 p_2^3 p_3^3\;.
\end{align}

In order to determine the statistics observed in sub-volumes, we split the Fourier expansion up into ``short" modes contained within a sub-volume and ``long" modes with wavelengths larger than the size of the sub-volume. The non-linear terms $Z_2$, $Z_3$, etc then contribute to lower order terms in the expansion for a ``short" mode when one or more of the integrated momenta are very long wavelength.

For example, when one of the momenta $p_i$ has a long wavelength, the $Z_2(\mathbf{k})$ term will contribute a shift to the linear piece of the field observed in a sub-volume, as written in Eq.\eqref{eq:expansion-chi-in-Z}. Using the $p_i\ll p_j\approx k$ limit of the kernel $N_2$ from Eq.(\ref{eq:N23}), the linear term in the expansion of the short wavelength mode is shifted to
\begin{align}
	\label{eq:zetashort}
	\zeta_{\text{NG}}(\mathbf{k})|_{\rm obs} &	= f_L(\mathbf{k})\zeta_G(\mathbf{k})+\dots \nonumber \\
& = \zeta_{\text{G}} (\mathbf{k}) \left[1 + C_{\text{NL}}^{(2)}(\nu) \left( \frac{k}{k_0} \right)^{\nu-3/2}\int_{k_{\text{IR}}}^{k_0} \frac{d^3p}{(2\pi)^3} \left( \frac{p}{k_0} \right)^{3/2-\nu} \zeta_{\text{G}}(\mathbf{p}) \right] + \dots \nonumber \\
	& = \zeta_{\text{G}} (\mathbf{k}) \left[1 + C_{\text{NL}}^{(2)}(\nu) \left( \frac{k}{k_0} \right)^{\nu-3/2} \zeta_L \right] + \dots
\end{align}
Here, 
\begin{equation}
	C_{\text{NL}}^{(2)}(\nu)= -f_{\text{NL}} \frac{3^{3/2} 2^{3\nu-1/2} \Gamma(\nu)}{4^{\nu}\pi N_{\nu}(8/27)}
	\label{eq:def-C2NL}
\end{equation}
collects the terms coming from the expansion of the Neumann function in Eq.~\eqref{eq:ansatz-bispectrum} as well as the normalization factors defined to recover the local shape ansatz in the equilateral limit. This expressions derived above are only correct for a sufficiently squeezed configuration of the bispectrum, $\frac{k_L}{k_S} \ll e^{-1/\nu}$ \cite{Chen2010a}, and so in particular one should not naively extrapolate the expressions above for $C_{\text{NL}}^{(2)}(\nu)$ for $\nu\ll 3/2$. However, since the only significant cosmic variance comes from the squeezed limit, the precise form of $C_{\text{NL}}^{(2)}(\nu)$ will not change the results we quote below. The value of $C_{\text{NL}}^{(2)}(\nu)$ is about $0.57f_{\text{NL}}$ for $\nu\approx 3/2$, and increases slowly (so that $C_{\text{NL}}^{(2)}(1.45)=0.6f_{\text{NL}}$, for example).

In writing Eq.(\ref{eq:zetashort}) also have defined the cumulative long wavelength background -- a constant for any particular sub-volume -- as
\begin{equation}
	\zeta_L \equiv \int_{k_{\text{IR}}}^{k_0} \frac{d^3p}{(2\pi)^3} \left( \frac{p}{k_0} \right)^{3/2-\nu} \zeta_{\text{G}}(\mathbf{p}).
	\label{eq:definition-zeta-long}
\end{equation}
An infrared cut-off, $k_{\text{IR}}$, is only needed for a sufficiently light field. The long wavelength background is a Gaussian field, assumed to be constant over patches of size $k_0^{-1}$, with mean zero and variance
\begin{equation}
	\langle\zeta^2_{L}(x)\rangle=\int_{k_{\text{IR}}}^{k_0} \frac{d^3p}{(2\pi)^3} \left(\frac{2\pi^2\Delta^2_{\zeta}}{p^3}\right)\left( \frac{p}{k_0} \right)^{3-2\nu}
\label{eq:variancephilong}
\end{equation}
Notice that $\langle\zeta^2_{L}(x)\rangle$ can only be substantially larger than $\Delta^2_{\zeta}$ when $\sigma$ is very light. Although the derivation of the bispectrum assumed a scale-invariant power spectrum, we can straightforwardly generalize this expression to allow for $n_s\neq1$. For light fields, this is
\begin{equation}
	\langle\zeta^2_{L}(x)\rangle=\int_{k_{\text{IR}}}^{k_0} \frac{d^3p}{(2\pi)^3} \left( \frac{p}{k_0} \right)^{2\varepsilon}P(p).
\label{eq:variance_ns}
\end{equation}
where
\be
\varepsilon=\frac{m^2}{3H^2}\;.
\label{eq:def_epsilon}
\ee
Confirmation that this is the correct generalization if we repeat the calculation of the quasi-single field bispectrum including leading order slow-roll corrections can be found in Appendix~\ref{app:spectralindex}.

From Eq.(\ref{eq:zetashort}) and Eq.(\ref{eq:variance_ns}) we can calculate how locally observed statistics vary about the mean of the large volume. The locally observed power spectrum, for example, can be shifted in both amplitude and scale-dependence from the power spectrum of the global volume (which is also the mean power spectrum, irrespective of sub-volume size):
\begin{equation}
	P^{\text{obs}}(k) \approx P_{\rm G}(k) \left[ 1 + 2C_{\text{NL}}^{(2)}\left( \frac{k}{k_0} \right)^{-\varepsilon} \zeta_L \right] + \dots
\label{eq:correction-P-from-B}
\end{equation}
where the subscript $G$ indicates that the power is the Gaussian power only (in the large volume) and the dots include inhomogeneous terms from coupling to long wavelength gradients as well as non-Gaussian corrections. (See Appendix B of \cite{Adhikari:2015yya} for a careful derivation of this expression.) Keep in mind that $C_{\text{NL}}^{(2)}$ depends on $\nu$. 

Performing the integral in Eq.(\ref{eq:variance_ns}) gives a better sense of how large the variations due to long wavelength modes can be:
\begin{equation}
	\langle\zeta^2_{L}(x)\rangle=\frac{\Delta^2_{\zeta}(k_0) [1-e^{-N_{\rm extra}(2\varepsilon +n_s-1)}]}{2\varepsilon+n_s-1}\;,
\label{eq:variance-of-long-mode}
\end{equation}
where $N_{\rm extra}\equiv \ln(k_0/k_{\text{IR}})$ is the number of e-folds of inflation before the mode $k_0$ exited the horizon (the ``extra" e-folds if $k_0$ is roughly the largest observable scale). Since $\Delta^2_{\zeta}(k_0)\ll1$, the variance Eq.~\eqref{eq:variance-of-long-mode} is small unless these two conditions are satisfied:
\begin{enumerate}
	\item There is sufficient power in long wavelength modes, meaning $n_s-1\lesssim0$, and
	\item There is sufficient coupling between long and short wavelength modes, which for quasi-single field inflation is equivalent to the requirement that the non-Gaussian field coupled to the inflaton is sufficiently light, $\varepsilon\ll1$. Notice that the value of $n_s(k_0)$ fixed by observation essentially determines how heavy the field can be before long wavelength modes are irrelevant. 
\end{enumerate}
For a sufficiently light field and a sufficiently red power spectrum, this result diverges as the size of the large volume goes to infinity (i.e. $N_{\text{extra}}\to \infty$ ). However, it is reasonable to restrict ourselves to volumes where $\zeta$ can be defined as a small fluctuation. Imposing $\langle\zeta^2_{L}(x)|_{\nu=3/2}\rangle\ll1$ constrains $N_{\text{extra}}\lesssim 350$, for example, if $\Delta^2$ is set to the {\it Planck} value and $n_s=0.96$ on all scales.

In general, the probability that locally observed quantities differ significantly from the global mean depends dramatically on the amplitude and shape of all correlations present in the non-Gaussian statistics in the large volume \cite{Nelson:2012sb,Nurmi:2013xv, LoVerde:2013xka}. However, to demonstrate how important cosmic variance can be even in a very simple case, consider the effect of the quadratic term ($Z_2$) only, with a very conservative restriction to weak non-Gaussianity in the large volume. That is, we require that the $\mathcal{O}((C^{(2)}_{NL})^2)$ contribution to the power spectrum in the large volume is small ($P_{\zeta, {\rm NG}}(k)\approx P_{\zeta, {\rm G}}(k)$), which requires $(C^{(2)}_{NL})^2\langle\zeta^2_{L}(x)\rangle\ll1$. This can be enforced even for $N_{\text{extra}}=350$ if $f_{\text{NL}}=1$. Note that this restriction simplifies the calculations, but is otherwise not required: the level of non-Gaussianity in the large volume can differ significantly from that in sub-volumes, especially when terms beyond quadratic order are allowed. 

\begin{figure}
\centering
	\begin{tikzpicture}
	    \tikzstyle{every node}=[font=\scriptsize]
		\node[font=\normalsize] at (-0.35,4) {$\frac{\left| \Delta r\right|}{r}$};
		\node[font=\normalsize] at (4.4,-0.25) {$\frac{\left|\Delta n_s\right|}{n_s}$};
		\node[anchor=south west,inner sep=0] at (0,0) {\includegraphics[width=0.5\textwidth]{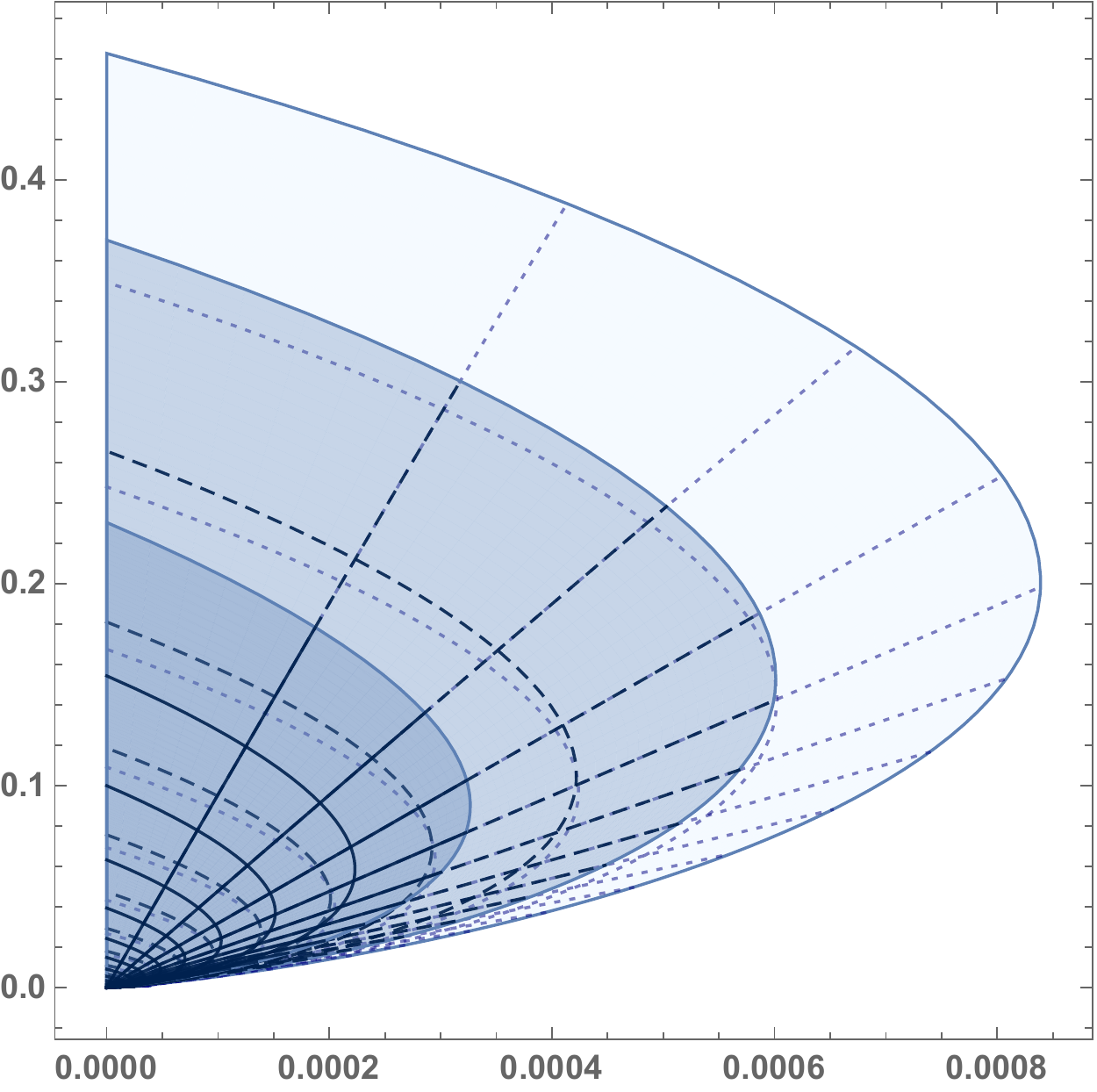}};
		\node[rotate=90] at (0.65,4) {$\nu = 1.5$};
		\node[rotate=18] at (5.5,2.53) {$\nu = 1.495$};
	\end{tikzpicture}
\caption{The figure shows the shift between the observed and global power spectrum parameters $\frac{ \left|\Delta r \right|}{r} = \frac{\left|r^{\text{obs}}-r\right|}{r}$ and $\frac{\left|\Delta n_s\right|}{n_s} = \frac{\left|n_s^{\text{obs}}-n_s\right|}{n_s}$ due to long and short modes coupling in a scenario where the large volume is weakly non-Gaussian (and the non-linear terms are only quadratic in the Gaussian field). The shift in parameters for sub-volumes one, two or three standard deviations from the mean value are shown (darkest to lightest regions respectively). The statistics of the large volume have $f_{\text{NL}}=1$. The radial lines of the mesh indicate lines of constant $\nu$ (from vertical line for $\nu=1.5$, i.e. $\frac{m}{H}=0$ to almost horizontal line for $\nu=1.46$, i.e. $\frac{m}{H}\simeq 0.34$), with a separation of $\Delta\nu=0.01$. The curved lines of the mesh are lines of constant $\langle\zeta_L^2\rangle$, or close to lines of constant $N_{\text{extra}}$ (since $\zeta_L$ depends on $n_s$, $\Delta_{\zeta}^2$ and $N_{\text{extra}}$, there is a degeneracy). The value $\nu=1.495$ corresponds to $\frac{m}{H}\simeq 0.12$. We set $k_0$ to be the largest CMB observable mode $k_0=0.008$ Mpc$^{-1}$, and values for the amplitude of the scalar spectrum and the spectral index at $k_*=0.05$ Mpc$^{-1}$ are taken from \cite{Ade:2015lrj}. Non-gaussian corrections to the power spectrum of the large volume have been taken into account to plot $\frac{\left| \Delta r\right|}{r}$, although they are not large in the scenario plotted here.}
\label{fig:rvsns}
\end{figure}
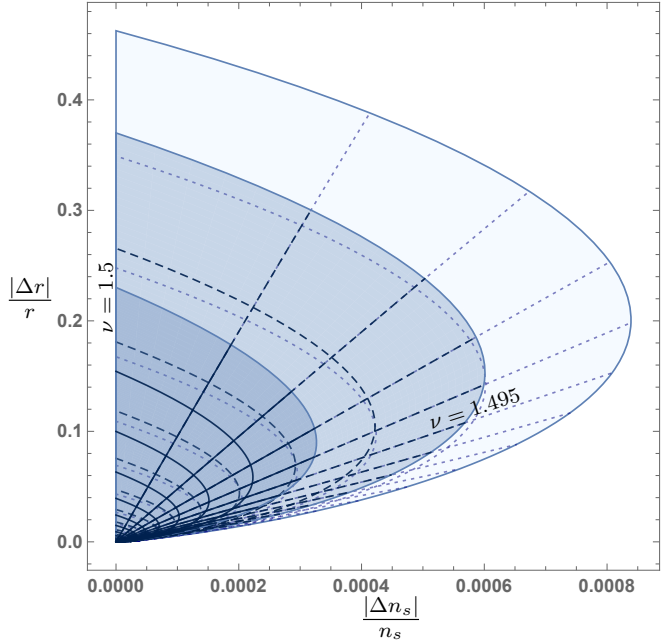

Then, for the simple scenario of large volume statistics given by a quadratic, weakly non-Gaussian expression, Eq.(\ref{eq:correction-P-from-B}) can be used to compute the shift in the observed amplitude and scale-dependence of the power spectrum. Assuming tensor modes are unaffected by cosmic variance, and that we insist the local volume amplitude of scalar fluctuations has the amplitude we observe, it is interesting to express the shift to the locally observed power spectrum amplitude as a change in the observed tensor-to-scalar ratio $r$. From Eq.(\ref{eq:correction-P-from-B}), this is just
\begin{align}
	\left. r^{\rm{obs}} \right|_{k=k_0}& \approx \frac{r}{\left[ 1 + 2C_{\text{NL}}^{(2)} \zeta_L \right]}.
\label{eq:r_obs}
\end{align}
The difference between the locally observed spectral index and the large volume (mean value) of $n_s$ is
\begin{align}
	\Delta n_s & = \left.n_s^{\text{obs}}\right|_{k=k_0}-n_s\nonumber\\
	&\simeq - \frac{2 \varepsilon \ C_{\text{NL}}^{(2)}  \zeta_L }{1+2 C_{\text{NL}}^{(2)} \zeta_L }.
\label{eq:shift_ns}
\end{align}
Although $\varepsilon$ grows with the mass of the field, $\zeta_L$ is strongly suppressed for large $\varepsilon$. 
Computing the bispectrum in the small volume by restoring the quadratic term for sub-volume modes (but using the shifted linear term), notice that
\begin{equation}
\fnl^{\rm obs}(k_0)\approx\frac{\fnl}{(1 + 2C_{\text{NL}}^{(2)} \zeta_L) }\;.
\end{equation}

Figure \ref{fig:rvsns} shows the fractional shift in the observed spectral index and tensor-to-scalar ratio ($\frac{\left|\Delta n_s \right|}{n_s}$ and $\frac{\left|\Delta r\right|}{r}=\frac{\left|r^{\text{obs}}-r \right|}{r}$) for sub-volumes that are one, two and three standard deviations from the mean (represented by the darker, medium and lighter blue zones, respectively). The meshes indicate lines of constant $\nu$ (from the vertical line for $\nu=1.5$, or $m=0$ to an almost horizontal line for $\nu=1.46$, corresponding to $m\simeq 0.34H$) as well as constant $\langle\zeta_L^2\rangle$, which are close to lines of constant $N_{\text{extra}}$ (there is a degeneracy due to the fact that $\langle\zeta_L^2\rangle$ depends on $\Delta_{\zeta}$, $n_s$ and $N_{\text{extra}}$). The outer boundary of the region is $N_{\text{extra}}=350$. As the figure shows, the shift to the spectral index is not significant unless the long-wavelength fluctuation is very rare, but the change in $r$ can be substantial. Notice that long wavelength backgrounds that have $\zeta_L,\,f_{\rm NL}>0$ enhance the local amplitude of fluctuations, which simultaneously suppresses the observed tensor-to-scalar ratio and the observed $f_{\rm NL}$. 

Finally, notice that using the observed amplitude of fluctuations and of the bispectrum at $k=k_0$ one would reconstruct a local, effective fluctuation Lagrangian with parameters shifted from their large volume values:
\begin{align}
\left[f(\nu)\left(\frac{\mu}{H} \right)\left(\frac{\rho}{H} \right)^3\right]^{\rm obs}&=(2 \pi \Delta_\zeta^{\rm obs})\fnl^{\rm obs} \nonumber \\
&\approx f(\nu)\left(\frac{\mu}{H} \right)\left(\frac{\rho}{H} \right)^3(1-2C_{\text{NL}}^{(2)}\zeta_L)\;.
\end{align}

\subsection{Comparison with the in-in calculation}
\label{sec:inin_split}
In the previous subsection, we analyzed the influence of the bispectrum on the squeezed limit on the power spectrum, in a classical, phenomenological calculation.
Here we reproduce the result via an approximate in-in computation.
Specifically, we show that the phenomenological result is equivalent to the correlation function obtained in the in-in calculation by comparing the expressions for the correction to the power spectrum due to all long wavelength modes $\langle \zeta_{\mathbf{k}_2} \zeta_{\mathbf{k}_3} \rangle_{\zeta_{L}}$, defined through the relation
\begin{equation}
	\langle \zeta_{\mathbf{k}_2} \zeta_{\mathbf{k}_3} \rangle^{\text{obs}} = \langle \zeta_{\mathbf{k}_2} \zeta_{\mathbf{k}_3} \rangle + \langle \zeta_{\mathbf{k}_2} \zeta_{\mathbf{k}_3} \rangle_{\zeta_{L}},
\end{equation}
in both formalisms.

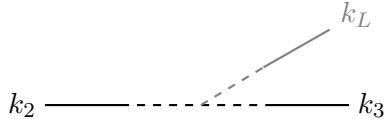
\begin{figure}
\begin{center}
\begin{tikzpicture}[scale=1]
	\draw [thick] (0,0) -- (1,0);
	\draw [thick, dashed] (1,0) -- (3,0);
	\draw [thick] (3,0) -- (4,0);
	\draw [thick, dashed, gray] (2.05,0) -- (2.875,0.5);
	\draw [thick,gray] (2.875,0.5) -- (3.745,1);
	\draw [gray] (4.1,1.2) node{$k_L$};
	\draw (-0.3,0) node{$k_2$};
	\draw (4.3,0) node{$k_3$};
\end{tikzpicture}
\end{center}
\caption{The vertex showing the correction to the power spectrum due to a long mode. The long mode, in grey, is outside the horizon and classical.}
\label{fig:theta-theta-with-long}
\end{figure}
We use the commutator form Eq.(\ref{eq:in-in-commutator}) for the in-in calculation of the vertex illustrated in \fref{fig:theta-theta-with-long} for a particular long wavelength mode $k_L$:
\begin{equation}
\begin{aligned}
	\langle \varphi_{\mathbf{k}_2} \varphi_{\mathbf{k}_3} \rangle_{\varphi_{\mathbf{k}_L}} & = i^4 \int^{\tau}_{-\infty} d \tau_1 \int^{\tau_1}_{-\infty} d \tau_2 \int^{\tau_2}_{-\infty} d \tau_3 \int^{\tau_3}_{-\infty} d \tau_4 \, \times \\
	& \quad \left\langle \left[H_{\text{int}}(\tau_4), \left[ H_{\text{int}} (\tau_3) , \left[ H_{\text{int}}(\tau_2) , \left[ H_{\text{int}} (\tau_1), \varphi_{\mathbf{k}_2}(\tau) \varphi_{\mathbf{k}_3}(\tau) \right]\right] \right] \right] \right\rangle.
\end{aligned}
\label{in-in-Hs}
\end{equation}
The interaction terms are:
\begin{align}
    H_{\text{SS}} (\tau) & = - \int d^3 x \ a^3 (\tau) \ \rho \ \varphi^{\prime}_{\text{S}} (\tau) \sigma_{\text{S}} (\tau) \\
    H_{\text{LL}} (\tau) & =  - \int d^3 x \ a^3 (\tau) \ \rho \ \varphi^{\prime}_{\text{L}} (\tau) \sigma_{\text{L}} (\tau)\\
    H_{\text{LSS}} (\tau) & = \int d^3 x \ a^4(\tau) \ \mu \ \sigma_{\text{L}} (\tau) \sigma_{\text{S}} (\tau) \sigma_{\text{S}} (\tau),
\end{align}
where the subscripts $S$ and $L$ designate the short and long modes respectively. Different permutations of these interaction vertices give rise to many terms in the full calculation but to get the characteristic behavior of such terms, it is sufficient to use the same strategy as in Section~\ref{sec:QSF} and reference \cite{Baumann2011b}.

Schematically, the contribution to the power spectrum from a long-wavelength mode in the bispectrum is
\begin{equation}
\begin{split}
	\langle \varphi_{k_{1,S}} \varphi_{k_{2,S}} \rangle_{\varphi_{k_L}} \sim\, &\delta^3(\kv_{1,S}+\kv_{2,S}+\kv_{L})\, (\varphi_{k_{1,S}} \, \varphi_{k_{2,S}})\bigr|_{\tau=0} \\
	&\times \prod^2_{i=1} \left(\int d \tau_i \; a^3\rho\varphi^{\prime}_{k_{i,S}} \sigma_{k_{i,S}}\right) \int d \tau a^3 \rho \varphi^{\prime}_{k_L} \sigma_{k_L}  \int d \tau a^4 \mu\sigma_{k_S}^{2}\sigma_{k_L}.
\end{split}
\label{eq:PS-schematic}
\end{equation}
The long mode can be dropped from the delta function and then the factor on the first line of Eq.(\ref{eq:PS-schematic}) is just $P_{\varphi}(k_{1,S})$. The last integral involving the cubic interaction vertex gives
\begin{equation}
\begin{split}
	\int d \tau a^4 \mu \sigma^2_{k_S} \sigma_{k_L}
	 \sim \frac{\mu}{H} \frac{k_S^{-3/2+\nu}}{k_L^{\nu}}.
\end{split}
\end{equation}
Each integral involving a transfer vertex of short wavelength modes gives rise to factors of $\rho/H$, as before. 

The same expression could be used for the interaction integral that depends only on the long wavelength mode, but note that the role of this term is conceptually different. When we are interested in the possible statistics in sub-volumes writing $\rho/H$ amounts to explicitly choosing a realization of $\varphi_{k_L}$ where the mode takes its r.m.s. value. To write the result in a way that clearly demonstrates that the long wavelength fluctuations should be drawn from a distribution, we instead keep $\varphi_{k_L}$ explicit. That is, to evaluate the integral we use $\varphi^{\prime}_{k_L} (\tau) \simeq k^2_L \tau \varphi_{k_L} (\tau)$. We take the $\tau\rightarrow0$ limit of the mode functions as before, but keep $\varphi_{k_L}$ an arbitrary constant (rather than replacing it by~$H/k_L^{3/2}$): 
 \begin{align}
	\int d\tau a^3 \rho \sigma_{k_L} \varphi^{\prime}_{k_L} & \sim  \int d\tau \frac{\rho}{H^2} (-\tau)^{-1/2-\nu} k_L^{-\nu +2} \varphi_{k_L}  \nn\\ 
	& \sim \frac{\rho}{H^2} k_L^{3/2} \varphi_{k_L}.
\end{align}
In considering this expression for the ensemble of Hubble patches sitting on top of long wavelength modes of size $k_L$, $\varphi_{k_L}$ is a stochastic variable drawn from a distribution with mean zero and variance $H^2/k_L^3$. Any single Hubble patch sits on a particular realization of the mode and $\varphi_{k_L}$ is just some constant. 

Putting everything together, converting the adiabatic perturbations to curvature perturbations, summing over all long wavelength modes, and relabeling momenta gives:
\begin{equation}
		\langle \zeta_{\mathbf{k}} \zeta_{\mathbf{k}^{\prime}} \rangle_{\zeta_{L}} \propto \left(2 \pi \right)^3 \delta \left( \mathbf{k} + \mathbf{k}^{\prime} \right)\frac{\Delta^2_{\zeta}}{k^3}\left[\left( \frac{\rho}{H} \right)^3    \left(\frac{\mu}{H} \right)  \frac{1}{\Delta_{\zeta}} \right] \left(\frac{k}{k_0}\right)^{\nu- \frac{3}{2}} \int_{k_{\text{IR}}}^{k_0} \frac{d^3 p}{(2\pi )^3} \left(\frac{p}{k_0} \right)^{\frac{3}{2}-\nu} \zeta_{\mathbf{p}},
\end{equation}
This agrees in all important qualitative features with the phenomenological expression Eq.(\ref{eq:correction-P-from-B}) where the term in square brackets has been replaced by $f_{\text{NL}}$ (up to numerical factors which should agree but cannot be checked with the approximately evaluated expressions used in this section).

\subsection{Variance of the mass of the hidden sector field}
\label{sec:shiftTS}

The next question is to determine whether long modes will affect the measurement of the mass of the hidden sector field, or in other words, the parameter $\nu$. The squeezed limit of the bispectrum, Eq.~\eqref{eq:bisp-in-squeezed-limit}, has a characteristic scaling $\sim (k_L/k_S)^{3/2-\nu}$, and we will show that the long mode coupling does not affect that scaling, regardless of the order of the self-interaction in the hidden sector.

From the limits of the correlation functions derived for quasi-single field inflation in the previous sections, we can determine the statistics of the power spectrum and squeezed limit bispectrum observed in sub-volumes. First, notice that the expansion in Eq.(\ref{eq:zng-expansion-in-Z}) generates the following expressions for the correlation functions:
\begin{equation}
	F_n(\kv_1,\kv_2,...\kv_n) \propto P(k_1)P(k_2)...P(k_{n-1})N_{n-1}(\kv_1,....,\kv_n)+(n-1\,\rm{perm.})
\end{equation}

When the correlation function is written as a single term, fully symmetrized over all momenta (as Eq.\eqref{eq:ansatz-bispectrum} and Eq.\eqref{eq:trispectrum-ansatz} are) the kernel is related to the correlation function by
\begin{equation}
	N_{n-1}(\kv_1,....,\kv_n)\propto\frac{1}{n}F^{\rm symm}_n(\kv_1,\kv_2,...\kv_n).
\end{equation}

To determine the shift to the squeezed limit of the bispectrum, we need to determine how the terms $Z_3$ and higher contribute to the local quadratic term. For that, we need the kernels $N_n$ in the limit where $n-2$ momenta correspond to super-horizon modes, and one mode is sub-horizon but still long compared to the remaining mode. The original quadratic kernel in the squeezed limit scales as
\be
N_2(p_1, p_2, k)\xrightarrow{p_1\ll p_2\approx k}\left(\frac{p_1}{k}\right)^{3/2-\nu}.
\ee
and more generally
\begin{equation}
	N_{n}(\pv_1,....,\pv_{n},k)\xrightarrow{p_1,p_2,\dots\ll p_n\approx k}\prod_{i=1}^{n-1}\left(\frac{p_{i}}{k}\right)^{3/2-\nu}\;.
\end{equation}
Using these kernels in the squeezed limit, the expression of the observed quadratic kernel, $Z_2^{\text{obs}}$ defined in Eq.\eqref{eq:Z2-obs-def}, can be expressed as a series:
\begin{equation}
\begin{split}
\fnl^{\rm obs}Z_2^{\rm obs, \,sqz.}(\kv)& = \frac{1}{2!(2\pi)^3}\int_{k_0<p_1\ll p_2} \hspace{-3.3em} d^3p_1\,d^3p_2 \; \delta^{(3)}\left(\mathbf{k}-\mathbf{p}_1 - \mathbf{p}_2 \right) \frac{\chi_G(\mathbf{p}_1)\chi_G(\mathbf{p}_2)}{f_L(\pv_1)f_L(\pv_2)}\,\\
 & \quad \times\left(\frac{p_1}{k}\right)^{\tfrac{3}{2}-\nu} \sum_{n=2}C^{(n)}_{\text{NL}} (\nu )\left( \left(\frac{k}{k_0}\right)^{\nu-\tfrac{3}{2}} \chi_L\right)^{n-2} 
\end{split}
\label{eq:Z2-obs}
\end{equation}
where $C^{(n)}_{\text{NL}}(\nu)$ is a function similar to Eq.\eqref{eq:def-C2NL}, and the long mode $\chi_L$ is just $\zeta_L$ expressed in terms of $\chi_G$:
\begin{equation*}
\label{eq:chi_long}
	\chi_L = \int_{k_{\text{IR}}}^{k_0} \frac{d^3p}{(2\pi)^3} \left( \frac{p}{k_0} \right)^{3/2-\nu} \frac{\chi_{\text{G}}(\mathbf{p})}{f_L(\pv)}.
\end{equation*}
The result in Eq.\eqref{eq:Z2-obs} is only valid in the limit where $p_1 \ll p_2 \sim k$. Keeping a generic expression of the kernels gives the same result, but the notation is cumbersome and makes the expressions less clear.

The shifted bispectrum in the squeezed limit $k_1 \ll k_2 \approx k_3$ is then:
\begin{align}
	\langle \zeta(\mathbf{k}_1) \zeta(\mathbf{k}_2) \zeta(\mathbf{k}_3) \rangle^{\text{obs}} & = \langle \chi(\mathbf{k}_1) \chi(\mathbf{k}_2) Z_2^{\text{obs}} (\mathbf{k}_3)\rangle + \text{permutations} \\
\begin{split}
	& \xrightarrow{k_1 \ll k_2 \approx k_3} \frac{1}{(2\pi)^3}\int_{p_{1,2}>k_0} \hspace{-2.3em} d^3p_1 d^3p_2 \, \langle \chi_G(\mathbf{k}_1)\chi_G(\mathbf{k}_2) \chi_G(\mathbf{p}_1)\chi_G(\mathbf{p}_2) \rangle \\
 	&  \quad \times \left(\frac{p_1}{k_3}\right)^{\tfrac{3}{2}-\nu} \sum_{n=2} C^{(n)}_{\text{NL}}(\nu)  \left( \left(\tfrac{k_3}{k_0}\right)^{\nu-\tfrac{3}{2}} \chi_L \right)^{n-2} \tfrac{ \delta^{(3)} \left(\mathbf{k}_3-\mathbf{p}_1 - \mathbf{p}_2 \right) }{f_L(\mathbf{p}_1)f_L(\mathbf{p}_2)}
 \end{split} \\
 \begin{split}
 	& = (2\pi)^3 \frac{ \delta^{(3)} (\mathbf{k}_1+\mathbf{k}_2+\mathbf{k}_3)}{f_L(\mathbf{k}_1)f_L(\mathbf{k}_3)} P^{\text{obs}}(k_1) P^{\text{obs}}(k_3) \left( \frac{k_1}{k_3} \right)^{\tfrac{3}{2}-\nu} \\
 	& \quad \times 2 \sum_{n=2}  C^{(n)}_{\text{NL}} \left( \left(\tfrac{k_3}{k_0}\right)^{\nu-3/2} \chi_L \right)^{n-2} 
 \end{split}
\end{align}
The last line above is independent of $k_1$. The amplitude of the bispectrum depends on $\chi_L$, but the \emph{power} of the ratio $k_L/k_S$ is $3/2-\nu$, which is the same power as in Eq.~\eqref{eq:bisp-in-squeezed-limit}. Therefore, the $\nu$ observed is not affected by long wavelength modes, and can still be read off from the scaling of the ratio between the long and short modes of the bispectrum in the squeezed limit. In other words, the mass of the hidden sector field can still be reliably extracted, without having to take into account the effects of long wavelength modes. This is in contrast to the case of the scalar spectral index whose shift is proportional to the realization of the long mode in the sub-volume.


\section{Discussion}
\label{sec:discussion}
We have investigated an inflationary scenario where a spectator scalar field is coupled to the inflaton in such a way that it can source significant non-Gaussianity in the curvature perturbations. We computed squeezed limits of the correlation functions from the contact diagrams of arbitrary non-derivative self-interactions of the spectator field. The resulting non-Gaussianity couples long and short wavelength modes to a degree that depends on the mass of the spectator fluctuations. Any degree of long-short mode coupling causes statistics in sub-volumes to differ from the global mean statistics by an amount that depends on the amplitude of fluctuations with wavelengths larger than sub-volume size. We computed the dependence of the locally observed power spectrum and of the squeezed limit of the locally observed bispectrum on the amplitude of long-wavelength fluctuations. 

We find that when the spectator scalar is sufficiently light compared to the Hubble scale ($m_{\sigma}\lesssim0.1H$) the variation of statistics in sub-volumes the size of our observed universe can be significant. When the hidden sector field is lighter and the number of e-folds before the largest observable mode today exited the horizon is larger, the probability of significant shifts is larger. A cubic self-interaction alone changes both the amplitude and scale-dependence of the locally observed power spectrum. Although the shift in the spectral index is small unless the background fluctuations are very rare, the amplitude of fluctuations can shift significantly even for very small levels of non-Gaussianity. Assuming tensor fluctuations have no cosmic variance of their own, this means that the observed tensor-to-scalar ratio can be enhanced or suppressed compared to the mean. Figure \ref{fig:rvsns} quantifies the probability of a significant shift. More generally, the contact diagrams from arbitrary non-derivative self-interactions in the spectator Lagrangian generate a bispectrum observed in biased sub-volumes whose squeezed limit depends on the mass of the hidden sector field. For example, a purely quadratic self-interaction for the hidden sector fluctuations would generate a bispectrum with the same squeezed limit as the bispectrum from a cubic self-interaction, but only in biased sub-volumes. The fact that soft momenta modes of these diagrams preserve the squeezed limit of the bispectrum means that they do not affect the value of the mass of the hidden field any small-volume observer would obtain from the bispectrum. These conclusions may be equally well obtained either from soft limits of the full in-in calculations or, perhaps more straightforwardly, by performing a long-short wavelength split in the late-time correlation functions. We leave the extension of the calculations and results presented here to include exchange diagrams, derivative self-interactions, and additional fields for future work.

It is important to note that these long wavelength effects are distinct from those that {\it do} affect the mass of a light scalar in de Sitter space. For example, the loop diagrams for a light scalar with a quartic self-interaction have a naive infrared dependence that can be re-summed into a correction to the mass of the field \cite{Starobinsky:1994bd, Petri:2008ig, Burgess:2009bs}. The mass of the light scalar is corrected everywhere, not just in some sub-volumes, and it is the shifted mass that will be observed in the squeezed limit of the bispectrum. The mass correction is a dynamical effect of field theory in (quasi) de Sitter space; the sub-sampling of non-Gaussian statistics in the post-inflationary universe is not sensitive to the dynamics that generated the primordial fluctuations.

Our results are a concrete example of an initial conditions problem for inflation: when cosmological data is only good enough to measure some properties of some correlation functions (eg, the bispectrum averaged over the full sky), that data only allows us to uniquely construct a Lagrangian for the fluctuations in our observed volume. The local Lagrangian is consistent with a larger set of possible ``global" Lagrangians sourcing additional inflation and which may share some features (here the mass of the spectator field) but differ in others (here the spectator field's potential). Of course, in the absence of cosmic variance this is also true, but there is a straightforward order by order map between the dimension of terms in the fluctuation Lagrangian and the order of correlation function measured. Cosmic variance from long-short mode coupling breaks that mapping. Making additional measurements such as any position-dependence of the observed bispectrum or a detection of a trispectrum would remove some degeneracies, but not all. It would be interesting to extend our results to other multi-field scenarios to understand more generally which properties of the particle content and dynamics can be determined regardless of cosmic variance and which cannot.

\paragraph{Acknowledgment:} We thank Niayesh Afshordi and Louis Leblond for helpful discussions. We also thank David Seery for comments on the draft. B.B. is supported by NSF grant PHY-1505411 and a Frymoyer Fellowship. S.B. is supported in part by NSF grant PHY-1307408. A.D. and S.S. are funded by NSF Award PHY-1417385. In addition, A.D. and S.S. thank the Perimeter Institute for hospitality while this work was being finished. This research was supported in part by Perimeter Institute for Theoretical Physics. Research at Perimeter Institute is supported by the Government of Canada through Industry Canada and by the Province of Ontario through the Ministry of Economic Development \& Innovation.

\newpage

\appendix
\section{Evaluating new diagrams from the quartic interaction}
\label{app:trisp-estimates}
Here we present some additional details of the calculation of the trispectrum, supporting the approximate expression, Eq.(\ref{eq:trispectrum-ansatz}), used in the body of the paper. We have made use of Mathematica (and some helpful routines available from \cite{VanEsch}) to compute the full trispectrum, but will just extract the scaling in some momentum configurations of interest, similarly to what was done in \cite{Chen2010a} for the bispectrum. Although we do not find this more detailed look at the calculation any more physically illuminating than what is presented in the main body of the text, it does confirm that the arguments used there hold. In addition, it is needed to derive an expression for the $\nu$-dependent coefficient that appears in the amplitude of the trispectrum if one is interested in more precise numerical results for the amplitude of the induced bispectrum. Finally, we also use this section to demonstrate in more detail why the long-short split calculations do not contain the kind of subtleties associated with loop diagrams to the power spectrum.

\subsection{The trispectrum}
In evaluating Eq.(\ref{eq:in-in-commutator}) for the four-point function of $\zeta$ there will be one $\sigma^4$ interaction term and four transfer vertices to connect $\sigma$ legs to $\varphi$ legs. (Converting $\varphi$ to $\zeta$ involves only multiplicative factors.) So, there are five time integrals and five commutators to be expanded. The result can be organized as four terms for the possible non-trivial placements of the $\sigma^4$ interaction. Not all terms will contribute to the dominant scaling in the squeezed limit, so we need only examine one that does. For example, placing the $\sigma^4$ interaction Hamiltonian at time $\tau_3$ gives the following contribution:
\begin{align}
\langle\zeta^4\rangle & \propto \frac{1}{k_1k_2k_3k_4}{\rm Re}\left\{\int_{-\infty}^{0} d\tau_1\cdots\int_{-\infty}^{\tau_4} d\tau_5\,(-\tau_1)^{-1/2}(-\tau_2)^{-1/2}(-\tau_3)^{2}(-\tau_4)^{-1/2}(-\tau_5)^{-1/2}\right. \nonumber \\
 &\times \sin(-k_4\tau_1)\sin(-k_3\tau_2)H_{\nu}^{(1)}(-k_4\tau_3)H_{\nu}^{(1)}(-k_3\tau_3)H_{\nu}^{(2)}(-k_4\tau_1)H_{\nu}^{(2)}(-k_3\tau_2)  \label{eq:trispect_first_line}
 \\
& \times\left. \left[H_{\nu}^{(2)}(-k_2\tau_3)H_{\nu}^{(1)}(-k_2\tau_4)e^{-ik_2\tau_4}-{\rm c.c.}\right] \left[H_{\nu}^{(2)}(-k_1\tau_3)H_{\nu}^{(1)}(-k_1\tau_5)e^{-ik_1\tau_5}-{\rm c.c.}\right] \right\}\; +\dots \nonumber
\end{align}
where the dots contain terms no more relevant than the one displayed. We want to extract dependence on the various momenta when one mode has very small momentum (the superhorizon mode) and the remaining modes are organized in a squeezed configuration. Choosing $k_4$ as the most UV mode (one of the short modes of the squeezed limit), we can write the above equation in terms of the dimensionless variables: $x_i=k_4\tau_i$. This yields
\begin{align}
\langle\zeta^4\rangle\propto&\frac{1}{k_4^6}\frac{1}{k_1 k_2 k_3}{\rm Re}\left\{\prod_{i=1}^5\int dx_i \;(-x_1)^{-1/2}(-x_2)^{-1/2}(-x_3)^{2}(-x_4)^{-1/2}(-x_5)^{-1/2}\right.\nonumber\\
&\times \sin(-x_1)\sin\left(-\frac{k_3}{k_4}x_2\right)H_{\nu}^{(1)}(-x_3)H_{\nu}^{(1)}\left(-\frac{k_3}{k_4}x_3\right)H_{\nu}^{(2)}(-x_1)H_{\nu}^{(2)}\left(-\frac{k_3}{k_4}x_2\right)\nonumber\\
&\times\left[H_{\nu}^{(2)}\left(-\frac{k_2}{k_4}x_3\right)H_{\nu}^{(1)}\left(-\frac{k_2}{k_4}x_4\right)e^{-i(k_2/k_4)x_4}-{\rm c.c.}\right]\nonumber\\
&\left.\times\left[H_{\nu}^{(2)}\left(-\frac{k_1}{k_4}x_3\right)H_{\nu}^{(1)}\left(-\frac{k_1}{k_4}x_5\right)e^{-i(k_1/k_4)x_5}-{\rm c.c.}\right] \right\} +\dots
\label{eq:termtrispectrum}
\end{align}
where the limits of integration are understood to follow from Eq.(\ref{eq:trispect_first_line}).

Next, consider $k_1\ll k_4$. In this case the first Hankel function on the last line can be expanded for small argument, $H_{\nu}^{(2)}(z)\propto iz^{-\nu}$. (As argued in \cite{Chen2010a} for the bispectrum, the Hankel functions in the second line will strongly suppress any contributions with large $x_3$, which is required in order for $(k_1/k_4)x_3\sim1$.) The remaining dependence on the ratio $k_1/k_4$ can be extracted by defining $y_5\equiv\frac{k_1}{k_4}x_5$. The result after these two steps is
\begin{align}
\langle\zeta^4\rangle\propto&\frac{1}{k_4^6}\frac{1}{k_1 k_2 k_3}\left(\frac{k_1}{k_4}\right)^{-1/2-\nu}\nonumber\\
&\times{\rm Re}\left\{\prod_{i=1}^4\int dx_i \int_{-\infty}^{\frac{k_1}{k_4}x_4} dy_5\;(-x_1)^{-1/2}(-x_2)^{-1/2}(-x_3)^{2-\nu}(-x_4)^{-1/2}(-y_5)^{-1/2}\right.\nonumber\\
&\times \sin(-x_1)\sin\left(-\frac{k_3}{k_4}x_2\right)H_{\nu}^{(1)}(-x_3)H_{\nu}^{(1)}\left(-\frac{k_3}{k_4}x_3\right)H_{\nu}^{(2)}(-x_1)H_{\nu}^{(2)}\left(-\frac{k_3}{k_4}x_2\right)\nonumber\\
&\times\left[H_{\nu}^{(2)}\left(-\frac{k_2}{k_4}x_3\right)H_{\nu}^{(1)}\left(-\frac{k_2}{k_4}x_4\right)e^{-i(k_2/k_4)x_4}-{\rm c.c.}\right]\nonumber\\
&\left.\times\left[iH_{\nu}^{(1)}\left(-y_5\right)e^{-iy_5}-{\rm c.c.}\right]\right\}\;.
\label{eq:y_5}
\end{align}
The full shape of the bispectrum induced in biased sub-volumes would come from evaluating the remaining integrals in all configurations of $\kv_2$, $\kv_3$, $\kv_4$. However, here we are particularly interested in whether or not the squeezed limit of the induced bispectrum has the same dependence on sub-horizon long and short modes. So, we can make the further simplification that $k_2\ll k_3\approx k_4$, expanding the first Hankel function in the third line, and redefining $y_4\equiv\frac{k_2}{k_4}x_4$. In addition, the delta function of momenta now enforces $k_3\approx k_4$, so the contribution to the four-point function in this limit goes as
\begin{align}
\langle\zeta^4\rangle\propto&\frac{1}{k_4^6}\frac{1}{k_1 k_2 k_3}\left(\frac{k_1}{k_4}\right)^{-1/2-\nu}\left(\frac{k_2}{k_4}\right)^{-1/2-\nu}\nonumber\\
&\times{\rm Re}\left\{\prod_{i=1}^3\int dx_i \int_{-\infty}^{\frac{k_2}{k_4}x_3} dy_4\int_{-\infty}^{\frac{k_1k_2}{k_4^2}y_4} dy_5\;(-x_1)^{-1/2}(-x_2)^{-1/2}(-x_3)^{2-2\nu}(-y_4)^{-1/2}(-y_5)^{-1/2}\right.\nonumber\\
&\times \sin(-x_1)\sin\left(-x_2\right)H_{\nu}^{(1)}(-x_3)H_{\nu}^{(1)}\left(-x_3\right)H_{\nu}^{(2)}(-x_1)H_{\nu}^{(2)}\left(-x_2\right)\nonumber\\
&\left.\times\left[H_{\nu}^{(1)}\left(-y_4\right)e^{-iy_4}+{\rm c.c.}\right] \left[H_{\nu}^{(1)}\left(-y_5\right)e^{-iy_5}+{\rm c.c.}\right]\right\}\;.
\label{eq:y_4}
\end{align}
Since the remaining Hankel functions suppress contributions from $|x_3|\gtrsim1$ and $k_2/k_4\ll1$, the upper limit of the $y_4$ integral can be taken to be 0, and similarly for the upper limit of the $y_5$ integral. Then, the bottom three lines of Eq.(\ref{eq:y_4}) just contribute a dimensionless number (but a function of $\nu$) and dependence on the momenta is
\begin{equation}
T\overset{k_1\ll k_2\ll k_3\approx k_4}{\longrightarrow}\frac{1}{(k_1 k_2)^{3/2+\nu}}\frac{1}{k_3^{6-2\nu}}\overset{\nu=3/2}{\rightarrow}\frac{1}{k_1^3k_2^3k_3^3}\;.
\end{equation}
The bispectrum induced in sub-volumes when $k_1$ is a super horizon mode can be recovered from the line above by multiplying by $k_1^3$ and replacing any remaining $k_1$ factors by the smallest observable wavelength $k_0$. This recovers the squeezed limit of the induced bispectrum (with $k_2\equiv k_L$, $k_3\equiv k_S$) used in the text:
\be
B^{\rm induced}(k_S,k_S,k_L)\propto \frac{1}{k_S^3k_L^3}\left(\frac{k_L}{k_S}\right)^{\frac{3}{2}-\nu}\left(\frac{k_0}{k_S}\right)^{\frac{3}{2}-\nu}.
\ee
Note that when $\nu=\frac{3}{2}$, this trispectrum will have the same effect on the locally observed bispectrum and power spectrum as the standard trispectrum from extending the local ansatz to cubic order:
\be
\Phi_{NG}(x)=\phi_G(x)+\fnl(\phi(x)^2-\langle\phi(x)^2\rangle)+\gnl\phi(x)^3.
\label{eq:local_ansatz_cubic}
\ee
The (exact) trispectrum from this ansatz is
\be
T(k_1,k_2,k_3,k_4)=\gnl P(k_1)P(k_2)P(k_3) + {\rm perm.}
\ee
which induces the usual local bispectrum in biased sub-volumes even if the quadratic term is absent from Eq.(\ref{eq:local_ansatz_cubic}). To see this, choose $k_1$ to be the super-horizon mode and multiply the previous equation by $k_1^3$. Then label $k_2$ the short sub-horizon mode and $k_3$ the long sub-horizon mode to find the contribution with the largest squeezed limit:
\be
B^{\rm induced}(k_S,k_S,k_L)\propto \frac{1}{k_S^3k_L^3}\;.
\ee

\subsection{The soft limits of the trispectrum versus the one loop diagram}
\label{sec:one-loop}
Naively, the double soft limit of the trispectrum from $\sigma^4$, which induces a shift to the power spectrum in biased sub-volumes, shares some features with the one-loop correction to the power spectrum. Both calculations contain integrals over long wavelength modes. Our results show that long wavelength background modes from the trispectrum do not affect the mass measured for the spectator field from the squeezed limit of the bispectrum. However, it is a classic result from the stochastic treatment of inflation that a scalar field with a quartic self-interaction receives a shift to its mass in a de Sitter background \cite{Starobinsky:1994bd}. This shift can be attributed to infrared effects and a signal of the shift is an apparent infrared divergence in the one-loop correction to the power spectrum. In this sub-section we review the one-loop calculation to clarify the difference between the dynamical infrared effect and sub-sampling.

The quartic interaction adds a correction to the equal time $\sigma$ field two-point function of the following form
\begin{align}
\langle\sigma(\kv_1, \tau)\sigma(\kv_2, \tau)\rangle_{\rm one\;loop}&\propto i\lambda\int_{-\infty}^{\tau} d\tau_1\, a^4(\tau_1)\int \dthree{q_1}\dthree{q_2}\dthree{q_3}\nonumber\\
&\times\langle\left[\sigma(\qv_1,\tau_1)\sigma(\qv_2,\tau_1)\sigma(\qv_3,\tau_1)\sigma(-\qv_1-\qv_2-\qv_3,\tau_1),\sigma(\kv_1,\tau)\sigma(\kv_2,\tau)\right]\rangle\nonumber\\
&\propto\delta^3(\kv_1+\kv_2)\, i\lambda\int_{-\infty}^{\tau} d\tau_1\,\frac{1}{\tau_1^4H^4}\int\dthree{q_3}\,v(\qv_3,\tau_1)v^*(\qv_3,\tau_1)\nonumber\\
&\times[v(\kv_1,\tau_1)v^*(\kv_1,\tau)v(\kv_2,\tau_1)v^*(\kv_2,\tau)-{\rm c.c.}]\;.
\label{eq:one-loop}
\end{align}
Aside from the difference in notation, our discussion of this equation follows the treatment in \cite{Burgess:2009bs}, where additional details may be found.

The remaining momentum integral in the second to last line of Eq.(\ref{eq:one-loop}) is divergent in both the UV and the IR for a massless field, and divergent in the UV for a massive field. The scale that appears in the upper limit after UV regularization (including appropriate counter terms in the full computation) and any IR scale imposed in the lower limit by the physics of the problem are both physical rather than co-moving scales. So, they enter the limits of integration together with the scale factor evaluated at the time the interaction takes place. (In the UV, the regularization scale is a physical scale because it should be insensitive to the curvature scale. In the IR,  convergence could come from the mass of the field or a fixed length scale when inflation began, both of which would provide a physical scale in the integral.) Expanding the mode functions to capture the leading infrared behavior, the momentum integral contributes a time-independent multiplicative factor in the 1-loop calculation:
\begin{equation}
\begin{split}
\int\dthree{q_3}\,v(\qv_3,\tau_1)v^*(\qv_3,\tau_1)&=-\frac{2^{2\nu}\Gamma(\nu)^2}{4\pi}\int_{a\Lambda_{IR}}^{a\mu}\dthree{q_3}H^2(\tau_1)^{-3}(-q_3\tau_1)^{-2\nu}\\
&=-\frac{2^{2\nu}\Gamma(\nu)^2}{4\pi(3-2\nu)}H^2\left[\left(\frac{\mu}{H}\right)^{3-2\nu}-\left(\frac{\Lambda_{IR}}{H}\right)^{3-2\nu}\right]
\end{split}
\end{equation}
where we have assumed $H$ to be time-independent. This expression holds for a massive field but the time-independence of the result is also true in the massless case. Clearly, this momentum integral is quite different from those that appear in the sub-sampling procedure (eg. Eq(\ref{eq:correction-P-from-B})): in that case the scale dividing the modes into ``long" and ``short" is arbitrary. The sub-sampling that breaks the integral into local and background pieces is done at a fixed time, after inflation ended (so in the $\tau\rightarrow0$ limit of any in-in calculation). There are no dynamical processes in the sub-sampling.

The interesting feature of the loop calculation can be seen by expanding the last line of Eq.(\ref{eq:one-loop}) to extract the IR limits of the mode functions. Using the leading imaginary and real parts of the Hankel function in the $(-k\tau_1\rightarrow0)$ limit and collecting the remaining $\tau_1$ dependence from the second line gives
\begin{align}
\langle\sigma(\kv_1, \tau)\sigma(\kv_2, \tau)\rangle_{\rm one\;loop}&\propto\langle\sigma(\kv_1, \tau)\sigma(\kv_2, \tau)\rangle_{\rm tree}\int_{-1/k_1}^{\tau}\frac{d\tau_1}{\tau_1}\left[\left(\frac{\tau_1}{\tau}\right)^{-2\nu}-1\right]
\end{align}
which diverges as $\ln(-k\tau_1)$ for $k\tau_1\rightarrow0$. This divergence appears at each order in the series of loop diagrams from the quartic interaction and can be re-summed into a shift to the mass of $\sigma$. Notice that in the calculation of the tree-level four-point for $\sigma$, the lack of the loop integral preserves the additional powers of $\tau_1$ from the mode functions in the interaction Hamiltonian. Those factors eliminate the divergence in the $\tau_1$ integral.

\section{Treating the exchange diagrams} 
\label{app:exchange}
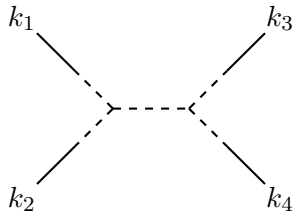
\begin{figure}[hb]
	\centering
	\begin{tikzpicture}
		\draw[thick, dashed] (0,0) -- (-.5,0.5);
		\draw[thick, dashed] (0,0) -- (-.5,-0.5);
		\draw[thick, dashed] (0,0) -- (1,0);
		\draw[thick, dashed] (1,0) -- (1.5,0.5);
		\draw[thick, dashed] (1,0) -- (1.5,-0.5);
		\draw[thick] (1.5,0.5) -- (2,1);
		\draw[thick] (1.5,-0.5) -- (2,-1);
		\draw[thick] (-0.5,0.5) -- (-1,1);
		\draw[thick] (-0.5,-0.5) -- (-1,-1);
		\draw (-1.2,1.2) node{$k_1$};
		\draw (-1.2,-1.2) node{$k_2$};
		\draw (2.2,1.2) node{$k_3$};
		\draw (2.2,-1.2) node{$k_4$};
	\end{tikzpicture}
\caption{The four-point interaction from the cubic interaction. As before solid lines represent the $\varphi$ field while dashed lines are $\sigma$.}
\label{fig:trisp_exchange}
\end{figure}

The cubic self interaction for the hidden sector field generates a four point function from two cubic vertices connected with an internal $\sigma$ field, shown in Figure \ref{fig:trisp_exchange}.

The in-in result for this exchange diagram is, schematically,
\begin{align}
	\langle \zeta_{\kv_1} \zeta_{\kv_2} \zeta_{\kv_3} \zeta_{\kv_4} \rangle_{\rm exchange}  \sim &\;(2 \pi)^3 \delta^3(\kv_1+\kv_2+\kv_3+\kv_4)\left(\zeta_{\mathbf{k}_1} \zeta_{\mathbf{k}_2} \zeta_{\mathbf{k}_3} \zeta_{\mathbf{k}_4} \right)|_{\tau= 0}\nn\\
	 & \times\left( \prod_{i=1}^{4} \int d\tau_i a^3 \rho \, \varphi^{\prime}_{k_i} \sigma_{k_i} \right)\nn\\
	 &\times \left( \int d\tau a^4 \mu \, \sigma_{k_1} \sigma_{k_2} \sigma_{| \mathbf{k}_1 + \mathbf{k}_2|} \right) \left( \int d\tau a^4 \mu \,  \sigma_{k_3} \sigma_{k_4}\sigma_{| \mathbf{k}_3 + \mathbf{k}_4 |} \right).
\label{eq:scalar_exchange}
\end{align}
As discussed in Section \ref{sec:kernels-trisp}, the integrals associated with the mixing term (on the second line above) contribute factors of $\rho / H$ and the multiplicative factor on the first line is just related to the power spectrum of $\zeta$. The remaining integrals over the $\sigma$ self-interactions, the last line of Eq.(\ref{eq:scalar_exchange}), depend on all momenta but can be approximately evaluated in configurations where one mode is significantly more UV than the others. The dominant contribution to these integrals comes when that mode finally crosses the horizon, $\tau\sim -1/k_{UV}$. 

\begin{figure}[hb]
\centering
	\begin{subfigure}{0.3\textwidth}
		\begin{tikzpicture}
			\draw[thick, dashed] (0,0) -- (-.5,0.5);
			\draw[thick, dashed] (0,0) -- (-.5,-0.5);
			\draw[thick, dashed] (0,0) -- (1,0);
			\draw[thick, dashed] (1,0) -- (1.5,0.5);
			\draw[thick, dashed] (1,0) -- (1.5,-0.5);
			\draw[thick] (1.5,0.5) -- (2,1);
			\draw[thick] (1.5,-0.5) -- (2,-1);
			\draw[thick] (-0.5,0.5) -- (-1,1);
			\draw[thick] (-0.5,-0.5) -- (-1,-1);
			\draw (-1.2,1.2) node{$k_1$};
			\draw (-1.2,-1.2) node{$k_2$};
			\draw (2.2,1.2) node{$k_3$};
			\draw (2.2,-1.2) node{$k_4$};
			\draw (0.5,0.3) node{$k_L$};
		\end{tikzpicture}
		\caption{Collapsed limit \\ \ \\ \ }
		\label{subfig:trisp_exchange_labelled_collapsed}
	\end{subfigure}
	\hspace{1em}
	\begin{subfigure}{0.3\textwidth}
		\begin{tikzpicture}
			\draw[thick, dashed] (0,0) -- (-.5,0.5);
			\draw[thick, dashed] (0,0) -- (-.5,-0.5);
			\draw[thick, dashed] (0,0) -- (1,0);
			\draw[thick, dashed] (1,0) -- (1.5,0.5);
			\draw[thick, dashed] (1,0) -- (1.5,-0.5);
			\draw[thick] (1.5,0.5) -- (2,1);
			\draw[thick] (1.5,-0.5) -- (2,-1);
			\draw[thick] (-0.5,0.5) -- (-1,1);
			\draw[thick] (-0.5,-0.5) -- (-1,-1);
			\draw (-1.2,1.2) node{$k_S$};
			\draw (-1.2,-1.2) node{$k_L$};
			\draw (2.2,1.2) node{$k_S$};
			\draw (2.2,-1.2) node{$k_L$};
		\end{tikzpicture}
	\caption{Long modes on opposite sides, connecting to different cubic vertex}
	\label{subfig:trisp_exchange_labelled_opposite}
	\end{subfigure}
	\hspace{1em}
	\begin{subfigure}{0.3\textwidth}
		\begin{tikzpicture}
			\draw[thick, dashed] (0,0) -- (-.5,0.5);
			\draw[thick, dashed] (0,0) -- (-.5,-0.5);
			\draw[thick, dashed] (0,0) -- (1,0);
			\draw[thick, dashed] (1,0) -- (1.5,0.5);
			\draw[thick, dashed] (1,0) -- (1.5,-0.5);
			\draw[thick] (1.5,0.5) -- (2,1);
			\draw[thick] (1.5,-0.5) -- (2,-1);
			\draw[thick] (-0.5,0.5) -- (-1,1);
			\draw[thick] (-0.5,-0.5) -- (-1,-1);
			\draw (-1.2,1.2) node{$k_S$};
			\draw (-1.2,-1.2) node{$k_S$};
			\draw (2.2,1.2) node{$k_L$};
			\draw (2.2,-1.2) node{$k_L$};
		\end{tikzpicture}
	\caption{Long modes on same side, connecting to the same cubic vertex}
	\label{subfig:trisp_exchange_labelled_same}
	\end{subfigure}
\label{fig:trisp_exchange_labelled}
\caption{The different configurations for which we evaluate the trispectrum. $k_S$ denotes a short mode, while $k_L$ denotes a long (either sub- or super-horizon) mode.}
\end{figure}
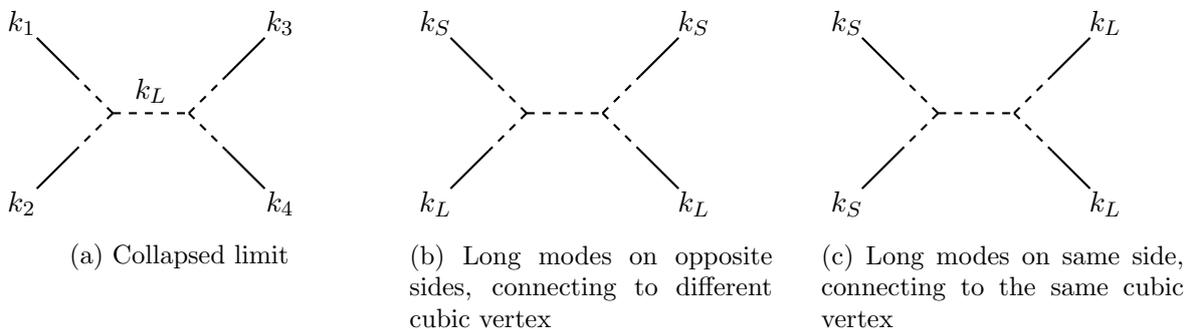

First, to see the difference in shape between the exchange diagram and the contact diagram, consider the ``collapsed" momentum configuration $k_{12} \equiv |\kv_1+\kv_2|\approx|\kv_3+\kv_4| \ll k_1\approx k_2 \approx k_3\approx k_4$, as shown on Fig.\eqref{subfig:trisp_exchange_labelled_collapsed}. In this case the trispectrum scales as
\begin{align}
\label{eq:exchangetrispectrum1}
	\lim_{\rm collapsed} T^{\,\rm exch.} _{\zeta}(\kv_1, \kv_2,\kv_3, \kv_4) &\propto \frac{\Delta^6_{\zeta}}{(k_1k_3)^{9/2-\nu}}\frac{1}{k_{12}^{2\nu}}\left( \frac{\mu}{H} \right)^2 \frac{1}{\Delta^2_{\zeta}}\left( \frac{\rho}{H} \right)^4 \\
	&\propto \frac{1}{\Delta^2_{\zeta}}\left( \frac{\mu}{H} \right)^2 \left( \frac{\rho}{H} \right)^4P(k_1)P(k_3)P(k_{12}) \left( \frac{k_{12}}{k_1} \right)^{\frac{3}{2}-\nu} \left( \frac{k_{12}}{k_3} \right)^{\frac{3}{2}-\nu}  \nonumber \\
	& \xrightarrow{\nu=3/2}\frac{1}{\Delta^2_{\zeta}}\left( \frac{\mu}{H} \right)^2 \left( \frac{\rho}{H} \right)^4P(k_1)P(k_3)P(k_{12})
\end{align}

Now consider the soft limits. The exchange diagram in the limit of two very soft momenta is sensitive to whether those modes are on the same side of the exchange diagram or not. In the limit $k_1\ll k_2\ll k_3\approx k_4$, as shown on Fig.\eqref{subfig:trisp_exchange_labelled_same}, the result is
\begin{align}
\label{eq:ex_tri_2soft}
\lim_{k_1\ll k_2\ll k_3\approx k_4} T^{\,\rm exch.} _{\zeta}(\kv_1, \kv_2,\kv_3, \kv_4) 
&\propto  \frac{\Delta^6_{\zeta}}{(k_1k_2k_3)^3}\left( \frac{\mu}{H} \right)^2 \frac{1}{\Delta^2_{\zeta}}\left( \frac{\rho}{H} \right)^4 \left(\frac{k_1}{k_3}\right)^{3/2-\nu}\\
& \propto \frac{1}{\Delta^2_{\zeta}}\left( \frac{\mu}{H} \right)^2 \left( \frac{\rho}{H} \right)^4P(k_1)P(k_2)P(k_3) \left(\frac{k_1}{k_3}\right)^{3/2-\nu} \nonumber \\
&\xrightarrow{\nu=3/2}\frac{1}{\Delta^2_{\zeta}}\left( \frac{\mu}{H} \right)^2 \left( \frac{\rho}{H} \right)^4P(k_1)P(k_2)P(k_3).
\end{align}
But, in the limit $k_1, k_3\ll k_2\approx k_4$, as depicted in Fig.\eqref{subfig:trisp_exchange_labelled_opposite}, the result is instead 
\begin{align}
\label{eq:ex_tri_2soft_opp}
\lim_{k_1, k_3\ll k_2\approx k_4} T^{\,\rm exch.} _{\zeta}(\kv_1, \kv_2,\kv_3, \kv_4) 
&\propto  \frac{\Delta^6_{\zeta}}{(k_1k_2k_3)^3}\left( \frac{\mu}{H} \right)^2 \frac{1}{\Delta^2_{\zeta}}\left( \frac{\rho}{H} \right)^4 \left(\frac{k_1}{k_2}\right)^{\frac{3}{2}-\nu}\left(\frac{k_3}{k_2}\right)^{\frac{3}{2}-\nu}\\
&\propto \frac{1}{\Delta^2_{\zeta}}\left( \frac{\mu}{H} \right)^2 \left( \frac{\rho}{H} \right)^4P(k_1)P(k_2)P(k_3) \left(\frac{k_1}{k_2}\right)^{\frac{3}{2}-\nu}\left(\frac{k_3}{k_2}\right)^{\frac{3}{2}-\nu} \nonumber \\
&\xrightarrow{\nu=3/2}\frac{1}{\Delta^2_{\zeta}}\left( \frac{\mu}{H} \right)^2 \left( \frac{\rho}{H} \right)^4P(k_1)P(k_2)P(k_3).
\end{align}

The momentum configurations studied above show that in the limit that $\sigma$ is massless, the exchange diagram generates a four-point function that is analogous to the $\tau_{\rm NL}$ familiar from the usual local ansatz. In the standard local ansatz, this shape is the trispectrum generated by the $f_{\rm NL}^2$ contribution from Eq.(\ref{eq:local_ansatz_cubic}). In that case, there is no additional kernel added at cubic order to generate this trispectrum, and no separate shift to power spectrum or bispectrum from the $\tau_{\rm NL}$ trispectrum. In the quasi-single field case, the quadratic kernel also generates a trispectrum. However, it scales as $\left( \frac{\mu}{H} \right)^2 \frac{1}{\Delta^2_{\zeta}}\left( \frac{\rho}{H} \right)^6$. The exchange diagram in Figure \ref{fig:trisp_exchange} is a significantly larger contribution by two factors of $\frac{H}{\rho}$. So for quasi-single field, we must add a new cubic kernel to account for this diagram (and subtract the sub-leading contribution from the quadratic kernel). The cubic kernel for the exchange diagram will be a generalization (to hold away from $\nu=3/2$) of the kernel for the $\tau_{\rm NL}$ trispectrum. However, to evaluate the effect of this new kernel on the power spectrum and squeezed limit of the bispectrum we need only the appropriate soft limits. To that end, notice that Eq.\eqref{eq:ex_tri_2soft_opp} has the same scaling behavior as we found for the contact diagram (see Eq.\eqref{eq:trisp-dsqueezed}, but note the difference in labels on the long wavelength momenta). So, this limit is already covered by the kernel used in the body of the paper. In the limit $\nu=3/2$, the scaling of Eq.(\ref{eq:ex_tri_2soft}) is also indistinguishable. We leave a more detailed analysis of this point for further work.

\section{The quasi-single field correlations when \texorpdfstring{$n_s\neq1$}{ns =/ 1}}
\label{app:spectralindex}
As we saw in Section \ref{sec:pheno-B-on-P}, long wavelength modes only have a significant effect on short wavelength correlations when the spectator field is light enough and when there is sufficient power in long wavelength modes. For a light field the growth of the quantity $\zeta_L$ depends on the difference of two small factors, $\varepsilon\approx\frac{m^2}{3H^2}$ and $n_s-1$, so it is important to be sure we have the right combination. In the text we used Eq.\eqref{eq:variancephilong}, originally derived assuming scale invariance,
\begin{equation}
	\langle\zeta^2_{L}(x)\rangle=\int_{k_{\text{IR}}}^{k_0} \frac{d^3p}{(2\pi)^3} \left( \frac{p}{k_0} \right)^{2\varepsilon}P(p).
\end{equation}
and then simply allowed the power spectrum to be scale-dependent. In this section we will verify that this is correct for quasi-single field inflation in an inflationary background with $n_s\neq1$.

The slow-roll parameters capture the deviation of the background evolution from that of an exact de Sitter space. We will work to leading order in slow-roll corrections, so we need only the first two parameters
\begin{align*}
\epsilon&\equiv-\frac{d\ln H}{dN}\\
\eta&\equiv-\frac{d\ln H_{,\Phi}}{dN}
\end{align*}
Where $N$ is the number of e-folds and $H_{,\Phi}$ is the derivative of the Hubble parameter with respect to the inflaton $\Phi$. There are several places where slow-roll parameters will correct the derivation of the bispectrum, trispectrum, and beyond. First, the equation satisfied by the mode functions is shifted so that the index of the Hankel functions also depends on slow-roll parameters. To leading order in slow-roll, for light fields
\be
\nu=\frac{3}{2}-\frac{1}{2}(n_s-1)-\frac{m^2}{3H^2}\;.
\label{eq:nu_ns}
\ee
In addition, the relationship between conformal time and scale factor is now
\be
aH\tau(1-\epsilon)=-1\;.
\ee

Repeating the qualitative derivation of the bispectrum (following \cite{Baumann2011b}), we need to evaluate the momentum dependence of
\begin{align}
 \langle \zeta_{\mathbf{k}_1} \zeta_{\mathbf{k}_2} \zeta_{\mathbf{k}_3} \bigr|_{\tau=0}  =  (2 \pi)^3 \left(\zeta_{\mathbf{k}_1} \zeta_{\mathbf{k}_2} \zeta_{\mathbf{k}_3} \right) \bigr|_{\tau=0} \left( \displaystyle \prod_{i=1}^{3} \int d\tau_i a^3 \rho \, \varphi^{\prime}_{k_i} \sigma_{k_i} \right) \left( \int d\tau a^4 \lambda \, \sigma_{k_1} \sigma_{k_2} \sigma_{k_3} \right).
    \label{eq:int-squeezed-thre-point}
\end{align}
in the squeezed limit $k_1\ll k_2\approx k_3$. As in the calculation of the trispectrum, the integrals inside the first set of parenthesis only contribute factors of $\rho/H$ with no momentum dependence. The last integral is dominated by the time when the most UV mode crosses the horizon and can be evaluated in the same manner as the interaction integral for the trispectrum in Section \ref{sec:kernels-trisp}.

The leading order slow-roll corrections to the momentum dependence will come from the factor out in front of the integrals and the dependence of $\nu$ on $n_s$. The fact that $H$ now depends on scale and the additional factors of $(1-\epsilon)$ from converting between scale factor and conformal time will give sub-leading corrections. Putting the pieces together, the scaling of the bispectrum is now
\be
 \langle \zeta_{\mathbf{k}_1} \zeta_{\mathbf{k}_2} \zeta_{\mathbf{k}_3} \rangle|_{k_1\ll k_2\approx k_3}\propto\frac{\Delta_{\zeta}^3}{k_1^3k_2^3}\left(\frac{k_1k_2^2}{k_0^3}\right)^{\frac{1}{2}(n_s -1)}\left(\frac{k_1}{k_2}\right)^{3/2-\nu}\;.
\ee
The ansatz for the bispectrum can be modified to include the appropriate powers of the spectral index:
\begin{align}
B(k_1,k_2,k_3) =B(k_1,k_2,k_3)|_{n_s=1}\left(\frac{k_1k_2k_3}{k_0^3}\right)^{\frac{1}{2}(n_s-1)}
\end{align}
and similarly for the kernel
\be
N_2(k_1,k_2,k_3)=N_2(k_1,k_2,k_3)|_{n_s=1}\left(\frac{k_3k_0}{k_1k_2}\right)^{\frac{1}{2}(n_s-1)}\;.
\ee
Finally, using this kernel to define the non-Gaussian field and performing the long-short split to determine the linear field in biased sub-volumes gives
\begin{align}
	\zeta_{\text{NG}}(\mathbf{k})|_{\rm obs} = \zeta_{\text{G}} (\mathbf{k}) \left[1 + C^{(2)}_{\text{NL}}(\nu) \left( \frac{k}{k_0} \right)^{\nu-3/2} \zeta_L \right] + \dots
\end{align}
where now
\begin{align}
	\zeta_L \equiv &\int_{k_{\text{IR}}}^{k_0} \frac{d^3p}{(2\pi)^3} \left( \frac{p}{k_0} \right)^{3/2-\nu} \left(\frac{p}{k_0}\right)^{-\frac{1}{2}(n_s-1)}\zeta_{\text{G}}(\mathbf{p})\nonumber\\
	=&\int_{k_{\text{IR}}}^{k_0} \frac{d^3p}{(2\pi)^3} \left( \frac{p}{k_0} \right)^{\varepsilon}\zeta_{\text{G}}(\mathbf{p})
\end{align}
where we have used Eq.(\ref{eq:nu_ns}) and $\varepsilon\approx\frac{m^2}{3H^2}$ as before.


\bibliographystyle{JHEP}

\begin{thebibliography}{10}

\bibitem{Ade:2015ava}
{\bf Planck} Collaboration, P.~A.~R. Ade et~al., {\it {Planck 2015 results.
  XVII. Constraints on primordial non-Gaussianity}},
  \href{http://arxiv.org/abs/1502.01592}{{\tt arXiv:1502.01592}}.

\bibitem{Alvarez:2014vva}
M.~Alvarez et~al., {\it {Testing Inflation with Large Scale Structure:
  Connecting Hopes with Reality}},  \href{http://arxiv.org/abs/1412.4671}{{\tt
  arXiv:1412.4671}}.

\bibitem{Linde:1996gt}
A.~D. Linde and V.~F. Mukhanov, {\it {Nongaussian isocurvature perturbations
  from inflation}},  {\em Phys. Rev. D} {\bf 56} (1997) 535--539,
  [\href{http://arxiv.org/abs/astro-ph/9610219}{{\tt astro-ph/9610219}}].

\bibitem{Moroi:2001ct}
T.~Moroi and T.~Takahashi, {\it {Effects of cosmological moduli fields on
  cosmic microwave background}},  {\em Phys.Lett.} {\bf B522} (2001) 215--221,
  [\href{http://arxiv.org/abs/hep-ph/0110096}{{\tt hep-ph/0110096}}].

\bibitem{Lyth2001}
D.~H. Lyth and D.~Wands, {\it {Generating the curvature perturbation without an
  inflaton}},  {\em Phys.Lett.} {\bf B524} (2002) 5--14,
  [\href{http://arxiv.org/abs/hep-ph/0110002}{{\tt hep-ph/0110002}}].

\bibitem{Enqvist:2001zp}
K.~Enqvist and M.~S. Sloth, {\it {Adiabatic CMB perturbations in pre - big bang
  string cosmology}},  {\em Nucl.Phys.} {\bf B626} (2002) 395--409,
  [\href{http://arxiv.org/abs/hep-ph/0109214}{{\tt hep-ph/0109214}}].

\bibitem{Dvali:2003em}
G.~Dvali, A.~Gruzinov, and M.~Zaldarriaga, {\it {A new mechanism for generating
  density perturbations from inflation}},  {\em Phys. Rev. D} {\bf 69} (2004)
  023505, [\href{http://arxiv.org/abs/astro-ph/0303591}{{\tt
  astro-ph/0303591}}].

\bibitem{Zaldarriaga:2003my}
M.~Zaldarriaga, {\it {Non-Gaussianities in models with a varying inflaton decay
  rate}},  {\em Phys. Rev. D} {\bf 69} (2004) 043508,
  [\href{http://arxiv.org/abs/astro-ph/0306006}{{\tt astro-ph/0306006}}].

\bibitem{Nelson:2012sb}
E.~Nelson and S.~Shandera, {\it {Statistical Naturalness and non-Gaussianity in
  a Finite Universe}},  {\em Phys. Rev. Lett.} {\bf 110} (2013), no.~13 131301,
  [\href{http://arxiv.org/abs/1212.4550}{{\tt arXiv:1212.4550}}].

\bibitem{Nurmi:2013xv}
S.~Nurmi, C.~T. Byrnes, and G.~Tasinato, {\it {A non-Gaussian landscape}},
  {\em JCAP} {\bf 1306} (2013) 004, [\href{http://arxiv.org/abs/1301.3128}{{\tt
  arXiv:1301.3128}}].

\bibitem{LoVerde:2013xka}
M.~LoVerde, E.~Nelson, and S.~Shandera, {\it {Non-Gaussian Mode Coupling and
  the Statistical Cosmological Principle}},  {\em JCAP 06,} {\bf 024} (2013)
  [\href{http://arxiv.org/abs/1303.3549}{{\tt arXiv:1303.3549}}].

\bibitem{LoVerde:2013dgp}
M.~LoVerde, {\it {Super cosmic variance from mode-coupling: A worked example}},
   {\em Phys. Rev. D} {\bf 89} (2014), no.~2 023505,
  [\href{http://arxiv.org/abs/1310.5739}{{\tt arXiv:1310.5739}}].

\bibitem{Byrnes:2013ysj}
C.~T. Byrnes, S.~Nurmi, G.~Tasinato, and D.~Wands, {\it {Implications of the
  Planck bispectrum constraints for the primordial trispectrum}},  {\em
  Europhys. Lett.} {\bf 103} (2013) 19001,
  [\href{http://arxiv.org/abs/1306.2370}{{\tt arXiv:1306.2370}}].

\bibitem{Dalal:2007cu}
N.~Dalal, O.~Dore, D.~Huterer, and A.~Shirokov, {\it {The imprints of
  primordial non-gaussianities on large-scale structure: scale dependent bias
  and abundance of virialized objects}},  {\em Phys. Rev. D} {\bf 77} (2008)
  123514, [\href{http://arxiv.org/abs/0710.4560}{{\tt arXiv:0710.4560}}].

\bibitem{Bartolo:2012sd}
N.~Bartolo, S.~Matarrese, M.~Peloso, and A.~Ricciardone, {\it {Anisotropic
  power spectrum and bispectrum in the $f(\phi)F^2$ mechanism}},  {\em Phys.
  Rev.} {\bf D87} (2013) 023504, [\href{http://arxiv.org/abs/1210.3257}{{\tt
  arXiv:1210.3257}}].

\bibitem{Thorsrud:2013mma}
M.~Thorsrud, D.~F. Mota, and F.~R. Urban, {\it {Local Observables in a
  Landscape of Infrared Gauge Modes}},  {\em Phys.Lett.} {\bf B733} (2014)
  140--143, [\href{http://arxiv.org/abs/1311.3302}{{\tt arXiv:1311.3302}}].

\bibitem{Thorsrud:2013kya}
M.~Thorsrud, F.~R. Urban, and D.~F. Mota, {\it {Statistics of Anisotropies in
  Inflation with Spectator Vector Fields}},  {\em JCAP} {\bf 1404} (2014) 010,
  [\href{http://arxiv.org/abs/1312.7491}{{\tt arXiv:1312.7491}}].

\bibitem{Bramante:2013}
J.~Bramante, J.~Kumar, E.~Nelson, and S.~Shandera, {\it {Cosmic variance of the
  spectral index from mode coupling}},  {\em JCAP} {\bf 2013} (2013), no.~11
  021--021, [\href{http://arxiv.org/abs/1307.5083}{{\tt arXiv:1307.5083}}].

\bibitem{Maldacena:2002vr}
J.~M. Maldacena, {\it {Non-Gaussian features of primordial fluctuations in
  single field inflationary models}},  {\em JHEP} {\bf 0305} (2003) 013,
  [\href{http://arxiv.org/abs/astro-ph/0210603}{{\tt astro-ph/0210603}}].

\bibitem{Creminelli:2004yq}
P.~Creminelli and M.~Zaldarriaga, {\it {Single field consistency relation for
  the 3-point function}},  {\em JCAP} {\bf 0410} (2004) 006,
  [\href{http://arxiv.org/abs/astro-ph/0407059}{{\tt astro-ph/0407059}}].

\bibitem{Hinterbichler:2013dpa}
K.~Hinterbichler, L.~Hui, and J.~Khoury, {\it {An Infinite Set of Ward
  Identities for Adiabatic Modes in Cosmology}},  {\em JCAP} {\bf 1401} (2014)
  039, [\href{http://arxiv.org/abs/1304.5527}{{\tt arXiv:1304.5527}}].

\bibitem{Joyce:2014aqa}
A.~Joyce, J.~Khoury, and M.~Simonovi\'c, {\it {Multiple Soft Limits of
  Cosmological Correlation Functions}},  {\em JCAP} {\bf 01} (2015) 012,
  [\href{http://arxiv.org/abs/1409.6318}{{\tt arXiv:1409.6318}}].

\bibitem{Mirbabayi:2014zpa}
M.~Mirbabayi and M.~Zaldarriaga, {\it {Double Soft Limits of Cosmological
  Correlations}},  {\em JCAP} {\bf 1503} (2015), no.~03 025,
  [\href{http://arxiv.org/abs/1409.6317}{{\tt arXiv:1409.6317}}].

\bibitem{2015PhRvD..91h3518B}
B.~Bayta{\c s}, A.~Kesavan, E.~Nelson, S.~Park, and S.~Shandera, {\it {Nonlocal
  bispectra from super cosmic variance}},  {\em Phys. Rev. D} {\bf 91} (Apr.,
  2015) 083518, [\href{http://arxiv.org/abs/1502.01009}{{\tt
  arXiv:1502.01009}}].

\bibitem{Chen:2009zp}
X.~Chen and Y.~Wang, {\it {Quasi-Single Field Inflation and
  Non-Gaussianities}},  {\em JCAP} {\bf 1004} (2010) 027,
  [\href{http://arxiv.org/abs/0911.3380}{{\tt arXiv:0911.3380}}].

\bibitem{Chen2010a}
X.~Chen and Y.~Wang, {\it {Quasi-single field inflation and
  non-Gaussianities}},  {\em JCAP} {\bf 04} (2010), no.~027 56,
  [\href{http://arxiv.org/abs/0911.3380}{{\tt arXiv:0911.3380}}].

\bibitem{Chen2010b}
X.~Chen and Y.~Wang, {\it {Large non-Gaussianities with intermediate shapes
  from quasi-single-field inflation}},  {\em Phys. Rev. D} {\bf 81} (2010),
  no.~6 1--5, [\href{http://arxiv.org/abs/0909.0496}{{\tt arXiv:0909.0496}}].

\bibitem{Sefusatti:2012ye}
E.~Sefusatti, J.~R. Fergusson, X.~Chen, and E.~P.~S. Shellard, {\it {Effects
  and Detectability of Quasi-Single Field Inflation in the Large-Scale
  Structure and Cosmic Microwave Background}},  {\em JCAP} {\bf 1208} (2012)
  033, [\href{http://arxiv.org/abs/1204.6318}{{\tt arXiv:1204.6318}}].

\bibitem{Norena:2012yi}
J.~Norena, L.~Verde, G.~Barenboim, and C.~Bosch, {\it {Prospects for
  constraining the shape of non-Gaussianity with the scale-dependent bias}},
  {\em JCAP} {\bf 1208} (2012) 019, [\href{http://arxiv.org/abs/1204.6324}{{\tt
  arXiv:1204.6324}}].

\bibitem{Baumann2011b}
D.~Baumann and D.~Green, {\it {Signature of supersymmetry from the early
  universe}},  {\em Phys. Rev. D} {\bf 85} (2012), no.~10 41,
  [\href{http://arxiv.org/abs/1109.0292}{{\tt arXiv:1109.0292}}].

\bibitem{Mirbabayi:2015hva}
M.~Mirbabayi and M.~Simonovi\'c, {\it {Effective Theory of Squeezed Correlation
  Functions}},  \href{http://arxiv.org/abs/1507.04755}{{\tt arXiv:1507.04755}}.

\bibitem{Arkani-Hamed:2015bza}
N.~Arkani-Hamed and J.~Maldacena, {\it {Cosmological Collider Physics}},
  \href{http://arxiv.org/abs/1503.08043}{{\tt arXiv:1503.08043}}.

\bibitem{Baumann2011}
D.~Baumann and D.~Green, {\it {Equilateral Non-Gaussianity and New Physics on
  the Horizon}},  {\em JCAP} {\bf 1109} (2011) 014,
  [\href{http://arxiv.org/abs/1102.5343}{{\tt arXiv:1102.5343}}].

\bibitem{Assassi2013}
V.~Assassi, D.~Baumann, D.~Green, and L.~McAllister, {\it {Planck-Suppressed
  Operators}},  {\em JCAP} {\bf 1401} (2014) 033,
  [\href{http://arxiv.org/abs/1304.5226}{{\tt arXiv:1304.5226}}].

\bibitem{Weinberg2005}
S.~Weinberg, {\it {Quantum contributions to cosmological correlations}},  {\em
  Phys. Rev. D,} {\bf 72} (Aug., 2005) 043514.

\bibitem{Adhikari:2015yya} 
  S.~Adhikari, S.~Shandera and A.~L.~Erickcek,
  Phys.\ Rev.\ D {\bf 93}, no. 2, 023524 (2016)
  doi:10.1103/PhysRevD.93.023524
  [arXiv:1508.06489 [astro-ph.CO]].

\bibitem{Ade:2013ydc}
{\bf Planck} Collaboration, P.~A.~R. Ade et~al., {\it {Planck 2013 Results.
  XXIV. Constraints on primordial non-Gaussianity}},  {\em Astron. Astrophys.}
  {\bf 571} (2014) A24, [\href{http://arxiv.org/abs/1303.5084}{{\tt
  arXiv:1303.5084}}].
  
\bibitem{Ade:2015lrj} 
{\bf Planck} Collaboration, P.~A.~R. Ade et~al., {\it {Planck 2015 results. XX. Constraints on inflation}}
  \href{http://arxiv.org/abs/1502.02114}{{\tt arXiv:1502.02114}}.
  
\bibitem{Starobinsky:1994bd}
A.~A. Starobinsky and J.~Yokoyama, {\it {Equilibrium state of a selfinteracting
  scalar field in the De Sitter background}},  {\em Phys. Rev. D} {\bf 50}
  (1994) 6357--6368, [\href{http://arxiv.org/abs/astro-ph/9407016}{{\tt
  astro-ph/9407016}}].

\bibitem{Petri:2008ig}
G.~Petri, {\it {A Diagrammatic Approach to Scalar Field Correlators during
  Inflation}},  \href{http://arxiv.org/abs/0810.3330}{{\tt arXiv:0810.3330}}.

\bibitem{Burgess:2009bs}
C.~P. Burgess, L.~Leblond, R.~Holman, and S.~Shandera, {\it {Super-Hubble de
  Sitter Fluctuations and the Dynamical RG}},  {\em JCAP} {\bf 1003} (2010)
  033, [\href{http://arxiv.org/abs/0912.1608}{{\tt arXiv:0912.1608}}].

\bibitem{VanEsch}
P.~Van~Esch, ``{Wick4.nb}.''
  \url{http://patrick.vanesch.pagesperso-orange.fr/qftcoursePS/wick4.nb}, 2002.
\newblock Accessed Feb 2014.

\end{thebibliography}
\providecommand{\href}[2]{#2}\begingroup\raggedright\endgroup
\end{document}